# Continual Model of Solvent: the DISOLV Software—Algorithms, Implementation, and Validation

O. Yu. Kupervasser[a], S. N. Zhabin[b], Ya. B. Martynov[c], K. M. Fedulov[d], I. V. Oferkin[e], A. V. Sulimov[f], and V. B. Sulimov[g]

Research Computer Center, Moscow State University

OOO Dimonta

[a] e-mail: e-mail: olegkup@yahoo.com
[b] e-mail: delafrog@gmail.com
[c] e-mail: yaroslavmartynov@yandex.ru
[d] e-mail: cf@dimonta.com
[e] e-mail: io@dimonta.com
[f] e-mail: sulimovv@mail.ru
[g] e-mail: vladimir.sulimov@gmail.com

**Abstract —** Several implicit (continuum) solvent models are considered: the Polarized Continuum Model (PCM), the Surface Generalized Born model (SGB), and the COnductor-like Screening MOdel (COSMO) as well as their implementation in the form of the DISOLV program. The methods for solving the corresponding equations and for computing the analytic gradients are described. The analytic gradients are used for the fast local energy optimization of molecules in a solvent. An algorithm for the original smooth triangulated molecular surface construction is shortly discussed. The procedure for matching the model parameters and the results of the program application to proteins and ligands with the employment of the MMFF94 force field are described. The validation results show the capability of the program to reach a good accuracy (about several tenth of kcal/mol) in the case of the solvation energy calculation for reasonable time periods at arbitrary shifts of the triangulated grid in use for such large molecules as proteins. A good agreement between the calculated and experimentally measured solvation energies in water is obtained with a root-mean-square deviation about 0.8 kcal/mol for several hundreds of molecules.

This study was performed as a part of the works of the Moscow State University on post-genome researches and technologies and the works on the program "Supercomputer Technologies for Solving the Problems of Processing, Storage, Transfer, and Protection of Information" (state contract no. 02.740.11.0388) and also supported in part by the Russian Foundation for Basic Research (projects nos. 09-01-12097_ofi-m and 10-07-00595-a).

**1. Introduction**

The main paradigm used in the modern development of new drugs consists in the following. Many diseases are associated with the functioning of certain proteins, so they must be blocked to cure diseases. For example, a target protein may belong to a virus, and its blockage disables the reproduction of a virus in an organism. Blockage is performed with the use of molecules that selectively bind to these proteins in an organism. Such molecules that serve as a basis for new drugs are called inhibitors. As a rule, inhibitors represent comparatively small organic molecules that bind to certain areas of target proteins. These areas are called binding sites or active sites. The search for such inhibitor molecules for a given target protein is the initial stage in the development of a new drug. The fast and efficient solution of this problem governs to a considerable degree the minimization of material expenditures and the duration of subsequent stages in the development of a new drug. With respect to time, the stage of developing new inhibitors takes nearly 50% of the total duration of the development of a new drug.

The time and material expenditures at the stage of searching for inhibitors can considerably be reduced with the use of computer-aided molecular modeling methods [1], among which docking is of first importance. Docking is the positioning of molecules that are candidates to inhibitors (they are often called ligands, from Latin *ligare* that means *to bind*) at the active site of a target protein and the estimation of their binding energy. The stronger a molecule binds to a protein, the better is an inhibitor and the more efficient is the drug based on this inhibitor. Docking is performed with the special molecular modeling software [2] that is also used on supercomputers.

The precision of estimating the protein–inhibitor binding energy governs the efficiency of predicting the activity of an inhibitor: the higher is the binding energy, the more active is an inhibitor and the more efficient is the drug based on this inhibitor, as the required effect can be attained at a lower drug concentration. If the precision of calculating the protein–inhibitor binding energy is insufficiently high, the probability of predicting whether new synthesized compounds (ligands) will inhibit this protein is low, and a great deal of materials, which are used to synthesize ligands and measure their inhibiting activity, and corresponding time are wasted. A sufficiently high practical predictability is attained at an error in calculating the protein–inhibitor binding energy of lower than 1 kcal/mol. For this reason, the precision of calculating all the contributions to the protein–inhibitor binding energy must be maximally high in the molecular modeling of the interaction of ligands with target proteins.

The precision of estimating the protein–inhibitor binding energy in molecular modeling is governed by many factors, e.g., the quality of a force field used for the description of intra- and intermolecular interactions, the efficiency of searching for a global minimum in the course of positioning an inhibitor in the active site of a target protein, the estimated contribution of the entropy component to the free protein–inhibitor binding energy, etc. Since the binding of an inhibitor to a protein in experiments (*in vitro* and *in vivo*) occurs in an aqueous solution, the presence of a solvent (water) must be taken into account, when calculating the protein–inhibitor binding energy.

The effect of a solvent on the protein–inhibitor binding energy is predominantly determined via the desolvation energy, which represents the difference between the solvation energy of a protein–ligand complex and the solvation energies of individual protein and ligand. This contribution to the protein–ligand binding energy is caused by that a solvent (water) is forced out of the active site of a protein upon the binding of a ligand to a protein, and some atoms in a ligand and a protein's active site cease to interact with a solvent. Hence, to determine the desolvation energy, it is necessary to calculate the solvation energies of a protein, a ligand, and their complex.

To calculate the free solvation energy, it is necessary to construct a solvent model. This may be done explicitly, considering a solvent as a set of a great number of molecules. However, this method needs comparatively high expenditures of computational resources in modeling, as the calculation of the observed effects requires us to perform the averaging over the state of solvent molecules, for example, by the molecular dynamics or Monte-Carlo methods. For this reason, implicit (often called continual) solvent models [3–5], in which a solvent is treated as a continuous (continual) medium with specified properties, including dielectric permittivity, are used more frequently.

The present work is devoted to the DISOLV software [6, 7] that allows us to find the free solvation energy of molecules and its gradients over the displacements of atoms in molecules, as the DISOLV software will be used to optimize the structure of molecules in a solvent with both the force field methods and the quantum-chemical methods, in which the gradient-based local optimization algorithms are usually applied.

The free Gibbs energy $\Delta G_s$ for the process of solvation, i.e., the transition of a molecule from a vacuum into a solvent or, briefly, the solvation energy, is represented as the sum of the three components

$\Delta G_s = \Delta G_{pol} + \Delta G_{np} + \Delta G_{cav}$,

where $\Delta G_{pol}$ is the polar component of the interaction of a dissolved substrate molecules with a solvent, $\Delta G_{np}$ is the non-polar component of the interaction of a dissolved molecule with a solvent due to van der Waals forces of intermolecular interaction, and $\Delta G_{cav}$ is the cavitation component of the free solvation energy due to the formation of a cavity comprising a dissolved molecule in the volume of a solvent.

In the DISOLV software, much attention is concentrated on the calculation of the polar component $\Delta G_{pol}$ of the interaction of a molecule with a solvent (or, briefly, the polarization energy) using several methods, and the other solvation energy components $\Delta G_{np}$ and $\Delta G_{cav}$ are taken into account by a simple and rather widely applied method (see below).

Within the framework of the used continual model, the solvation energy represents the energy of the electrostatic interaction of atomic charges in a molecule located in the cavity of a dielectric with surface charges induced by them on this surface.

By now, many existing implicit solvent models and their software implementations have been integrated into larger packages, e.g., the quantum-chemical packages Gaussian [8], Gamess [9], MolPro [10], MOPAC [11], the molecular dynamics package Charmm [12], and the free programs for finding the polar component of the interaction with a solvent, e.g., DelPhi [13] or APBS [14], implement the numerical solution of the finite-difference approximation of the three-dimensional Poisson–Boltzmann equation.

The objective of the present paper is to describe the original algorithms that are designed to calculate the polar component of the solvation energy of molecules and aimed at solving the corresponding equations on the two-dimensional surface of a solvent surrounding a molecule, their software implementation DISOLV written in C++, and its corresponding validation. At the end of the paper, we perform the brief comparison of the characteristics of DISOLV with the corresponding characteristics of APBS [14] that implements the numerical solution of the Poisson–Boltzmann equation in a 3D space.

**2. Physical Models**

In the continual model approximation, the molecule–solvent electrostatic interaction described by the summand $\Delta G_{pol}$ is estimated as the interaction of point charges located inside a cavity that is cut in the volume of a homogeneous dielectric or conductor (see below) with polarization charges that are induced on the internal surface of this cavity. A dielectric or a conductor acts as a solvent. The shape of a cavity that is cut in it corresponds to the shape of a molecule within an accuracy of the size of a solvent molecule. The point charges in a cavity describe a non-uniform distribution of electron density inside a molecule. Charges are usually placed in the centers of atoms that form a dissolved molecule. The charges are found from quantum-chemical calculations by replacing an electron distributed charge by a set of equivalent point charges or within the framework of force field, i.e., using the set of classic potentials that describe the interaction between atoms, when they are determined by the corresponding typization of atoms. In the present work, the atomic charges in molecules are determined by the Merck molecular force field (MMFF94) [15].

However, to solve the electrostatic problem, it is necessary to define beforehand, what is the surface of a molecule. There exist the two definitions of surfaces surrounding a molecule [16, 17]. The first of them is called SES (solvent excluded surface): the volume occupied by a solvent lies outside the volume enveloped by this surface, and the substrate itself lies completely inside this volume. The second is denoted as SAS (solvent accessible surface) and is formed by the centers of solvent molecules tangent to a substrate molecule.

SES corresponds to the boundary between the areas of a dissolved molecule and a solvent, i.e., it represents the surface of contact between the atoms of a dissolved substrate molecule and solvent molecules. This is precisely the surface, on which the charges induced in a dielectric must be located reasoning from their physical meaning. For this reason, SES is used just for electrostatic calculations.

In contrast to SES, SAS describes the boundary, near which the average concentration of solvent molecules varies from zero to the value that is average over the volume of a solvent. Since the non-electrostatic energy component is described by the number of solvent molecules tangent to the surface of a substrate, this number is proportional to the SAS area, whereupon the non-electrostatic energy component is often described by the linear equation in terms of the SAS area [3]. In the DISOLV software, both SES and SAS are constructed using the modified TAGSS version [18–20].

The DISOLV software employs the following four methods for calculating the polar component of the interaction of a dissolved substrate molecule with a solvent in the continual approximation.

### 2.1. Polarized Continuum Model (PCM)

This method represents a precise solution of the electrostatic problem of determining the energy of interaction between point charges and an ambient homogeneous continuous dielectric with a specified dielectric permittivity $\varepsilon$. In this method, the solution of the Poisson equation in a three-dimensional space is reduced to finding the surface charges induced by a dissolved molecule on the internal surface of a dielectric cavity that contains a molecule. In this method, instead of solving the Poisson equation in a three-dimensional space, we need to solve the corresponding equations on SES that bounds a solvent. We assume that the dielectric permittivity is $\varepsilon_{in} = 1$ within this surface and equal to $\varepsilon$ outside this surface. Then the polar component of the substrate–solvent interaction energy is determined as

$$\Delta G_{pol} = \frac{1}{2} \sum_i Q_i \int_{SES} \frac{\sigma(\vec{r})}{|\vec{R}_i - \vec{r}|} dS, \qquad (1)$$

where $\sigma(\vec{r})$ is the density of a polarization charge at the point $\vec{r}$ on SES, $\vec{R}_i$ is the vector determining the position of the charge $Q_i$ of each atoms in a substrate molecule, the integral is taken over SES, and summation is performed over all the charges of substrate atoms. The density of surface charges $\sigma(\vec{r})$ is found from the PCM equation [3, 21, 22] as

$$\sigma(\vec{r}) = \frac{1-\varepsilon}{2\pi(1+\varepsilon)} \left( \sum_i \frac{Q_i\left((\vec{r}-\vec{R}_i)\cdot\vec{n}\right)}{|\vec{r}-\vec{R}_i|^3} + \int_{SES} \frac{\sigma(\vec{r}\,')\left((\vec{r}-\vec{r}\,')\cdot\vec{n}\right)}{|\vec{r}-\vec{r}\,'|^3} dS' \right).$$

(2)

To find the numerical solution of Eq. (2), SES is divided into small elements, and Eq. (2) itself is represented in the matrix form as

$$Aq = BQ, \qquad (3)$$

where $A = \{a_{ij}\}$ is a square $N \times N$ matrix that depends on the parameters of surface elements, $N$ is the number of surface elements, $B = \{b_{ij}\}$ is an $N \times N$ matrix that depends on the geometric parameters of surface elements and the position of the charges $Q_i$ of atoms in a molecule, $M$ is the number of atoms in a molecule, $q = \{q_i = \sigma_i S_i\}$ is the column vector of the charges of surface elements, $\sigma_i$ is the average density of the $i^{th}$ surface element's charge, $S_i$ is the $i^{th}$ surface element's area, and $Q = \{Q_i\}$ is the column vector of the charges of atoms in a molecule. In contrast to Eq. (2), system (3) operates with the charges of surface elements instead of the surface density of a charge.

The elements of the matrix $B$ have the form

$$b_{ij} = \frac{\vec{n}_i (\vec{r}_i - \vec{R}_j)}{|\vec{r}_i - \vec{R}_j|^3} S_i \frac{1-\varepsilon}{4\pi(\varepsilon+1)}, \qquad (4)$$

where $\vec{r}_i$ is the vector determining the position of the $i^{th}$ surface element's center, $n_i$ is the vector of the normal at the $i^{th}$ surface element's center towards a dielectric solvent, and $\vec{R}_j$ is the vector determining the position of the $j^{th}$ surface element's center. With an accuracy of the coefficient $\frac{1-\varepsilon}{4\pi(\varepsilon+1)}$, each element of the matrix $B$ is the solid angle, at which the $i^{th}$ surface element can be seen from the $j^{th}$ atom.

For the sum of elements in each column of the matrix $B$, the equality used in DISOLV for the correction of numerical errors

$$\sum_{i=1}^{N} b_{ij} = \frac{1-\varepsilon}{1+\varepsilon}, j = 1, \ldots, M \qquad (5)$$

is true.

The elements of the matrix $A$ have the form [22]

$$a_{ij} = \frac{\vec{n}_i (\vec{r}_i - \vec{r}_j)}{|\vec{r}_i - \vec{r}_j|^3} S_i \frac{\varepsilon - 1}{4\pi(\varepsilon+1)} \text{ at } i \neq j, \ a_{jj} = \frac{\varepsilon}{\varepsilon+1} - \sum_{i \neq j} a_{ij}. \qquad (6)$$

In Eq. (6), the diagonal elements are determined from the normalization condition

$$\sum_i a_{ij} = \frac{\varepsilon}{\varepsilon+1}, j = 1, \ldots, N, \tag{7}$$

which follows from the equality $\int_{SES}\int_{SES} \frac{((\vec{r}-\vec{r}')\cdot\vec{n})}{|\vec{r}-\vec{r}'|^3} dS'dS = 2\pi S_{SES}$, where $S_{SES}$ is the total surface area [22]. The total charge induced on the surface and the total charge of all the atoms in a molecule are related as

$$\sum_{j=1}^{N} q_j = -\left(1-\frac{1}{\varepsilon}\right)\sum_{i=1}^{M} Q_i. \tag{8}$$

The expression for the energy (1) can be written as

$$\Delta G_{pol} = \frac{1}{2} Q^T D q, \tag{9}$$

where the elements of the matrix $D$ are written as

$$d_{ij} = \frac{1}{|\vec{R}_i - \vec{r}_j|}. \tag{10}$$

**2.2. Conductor-Like Screening Model (COSMO)**

For high $\varepsilon \gg 1$ as, for example, in the case of water with $\varepsilon = 78.5$, we can use the COSMO method [23]. In this model, a solvent is replaced by a metal or a dielectric with an infinite dielectric permittivity, i.e., $\varepsilon \to \infty$. After the surface charge induced by substrate charges is found, the corresponding energy is calculated by Eq. (1), and the obtained value is multiplied by the correcting factor $C_f$

$$C_f = \frac{\varepsilon - 1}{\varepsilon + 1/2}. \tag{11}$$

The relative error of COSMO has a value of the order of $\frac{1}{2\varepsilon}$. An advantage of COSMO over PCM consists in that the column of surface charges in the linear matrix equation is multiplied by a symmetric positive-definite matrix, whose elements do not depend on the normals of surface elements. For such a matrix, the energy and the analytical gradients can be found more quickly and precisely. The COSMO equation for the polarization charge in the integral form [23] is written as

$$\int_{SES} \frac{\sigma(\vec{r}')}{|\vec{r}-\vec{r}'|} dS' + \sum_{i=1}^{N} \frac{Q_i}{|\vec{r}-\vec{R}_i|} = 0,$$

where $r$ is the radius vector of any point on the surface or outside it. After the discretization of SES, i.e., its division into finite elements, this equation is written as

$$A^c q = -D^T Q, \tag{12}$$

where the matrix $D$ is determined by Eq. (10), the index $T$ means transposition, and the elements of the symmetric matrix $A^c$ have the form

$$a_{ij}^c = \frac{1}{|\vec{r}_i - \vec{r}_j|} \text{ at } i \neq j, \; a_{ij}^c = \frac{2\sqrt{3.83}}{\sqrt{S_i}}. \tag{13}$$

Note that Eq. (12) takes the form of Eq. (3), i.e., the same as in the PCM method, if the matrix $B$ with the elements $b_{ij} = -d_{ij}$ is introduced.

The polarization energy in the matrix form is written as

$$\Delta G_{pol} = \frac{1}{2} C_f Q^T D q. \tag{14}$$

Using the Ostrogradskii–Gauss theorem, we obtain the identity

$$\sum_{j=1}^{N} q_j = -\sum_{i=1}^{M} Q_i,$$

where $N$ is the number of surface elements, and $M$ is the number of charges inside a cavity.

**2.3. Surface Generalized Born (SGB) Model**

In the context of the heuristic SGB method, the polarization energy is expressed as [24]

$$\Delta G_{pol} = -\frac{1}{2}\left(1 - \frac{1}{\varepsilon}\right) \sum_{i,j} \frac{Q_i Q_j}{\sqrt{|\vec{R}_{i,j}|^2 + a_i a_j \cdot \exp\left(-\frac{|\vec{R}_{i,j}|^2}{c a_i a_j}\right)}}, \tag{15}$$

where $\vec{R}_{i,j} = \vec{R}_i - \vec{R}_j$, $c$ is an empirical constant equal to 8, and the Born radii $a_i$ are found via the integrals over SES as

$$a_i = \frac{1}{2}\left(\sum_{n=4}^{7} A_n I_n^i - A_0\right)^{-1}. \tag{16}$$

Here $A_n$ are empirical constants determined in [24], and $I_n^i$ are the integrals over SES:

$$I_n^i = \left( \oint \frac{(\vec{n}_s \cdot (\vec{r}_s - \vec{R}_i)) dS}{|\vec{r}_s - \vec{R}_i|^n} \right)^{1/(n-3)}, \quad 7 \geq n \geq 4. \tag{17}$$

**2.4. Polarized Continuum Model with Enlarged Surface Elements**

The PCM method with enlarged surface elements [22] is described by the same initial integral equations (1) and (2) as the conventional PCM method, but, in the course of discretization, small surface elements are combined into groups of elements that are closest to the given surface atom. We assume that the surface charge density is equal at all points for any enlarged surface element. Then the equation of PCM with enlarged surface elements in the matrix form is written as

$$Rq^{big} = EQ,$$

where $q^{big}$ is the column of the charges of enlarged surface elements.

The matrix elements for enlarged elements will be calculated by the formulas

$$R_{jk} = \frac{\sum_{l_j} \sum_{m_k} a_{l_j m_k} S_{m_k}}{\sum_{m_k} S_{m_k}} \text{ at } j \neq k, \quad R_{kk} = \frac{\varepsilon}{1+\varepsilon} - \sum_{j \neq k} R_{jk}, \quad E_{ji} = \sum_{l_j} b_{l_j i},$$

where $l_j$ and $m_k$ are the groups of small surface elements combined into a single big surface element, $S_{m_k}$ is the area of small surface elements, and $a_{l_j m_k}$ and $b_{l_j i}$ are the matrix elements determined for small surface elements by Eqs. (4) and (6). For $E_{ji}$, the normalization condition that is similar to (5)

$$\sum_j E_{ji} = \frac{1-\varepsilon}{1+\varepsilon}, \quad i = 1,...,M$$

is true.

**2.5. Calculating the Non-Polar Components of the Solvation Energy**

The non-polar components of the solvation energy $\Delta G_{np} + \Delta G_{cav}$ are calculated using the widely applied formula [25]

$$\Delta G_{np} + \Delta G_{cav} = \sigma S_{SAS} + b,$$

where $S_{SAS}$ is the SAS area. Note that SAS is formed by the centers of solvent molecules tangent to a dissolved substrate molecule. In the continual solvent model, the non-electrostatic component of the substrate–solvent interaction energy is assumed to be proportional to the number of solvent molecules tangent to a substrate molecule, and the proportionality coefficient determines the force of the van der Waals interaction between solvent molecules and a substrate. This coefficient and the correction constant implicitly take into account the cavitation energy, i.e., the free energy necessary to create a cavity, into which a substrate molecule is placed, in a homogeneous solvent. SAS exceeds SES by the occupied volume for the same substrate molecule and can be obtained from the latter via similarity transformation.

In [25], the parameters $\sigma = 0.00387$ kcal/(mol Å$^2$) and $b = 0.698$ kcal/mol are used for water. Since $\varepsilon_{in} = 1$ was used in DISOLV instead of $\varepsilon_{in} = 2.24$ in [25], these parameters were involved into the process of optimization along with the radii of atoms (see part 9) and, as a result, we obtained the same values for the non-polar component as in [25], namely, $\sigma = 0.00387$ kcal/(mol Å$^2$) and $b = 0.698$ kcal/mol.

### 3. Method of the Formation of SES and SAS

The DISOLV software employs the modified TAGSS (triangulated adaptive grid smooth surface) molecular surface formation subroutine [18–20], which performs the two main stages of constructing a surface: the first stage is primary and secondary rolling [16, 26], and the second stage is the generation of a triangulation grid using the parameters obtained at the first stage. A molecule is represented as a set of hard spheres, whose centers are located in the nuclei of atoms in a molecule, and the radii of these spheres are various for different types of atoms and are the parameters of the continual solvent model. Primary rolling is the formation of a molecular surface by rolling a molecule with a probe ball that simulates a solvent molecule. The two surfaces are constructed over a molecule: SAS described by the positions of the center of a rolling probe ball and SES composed by the points that are reached by a rolling probe ball and located most closely to the atoms of a rolled molecule (Fig. 1). In Fig. 1, $R_{pr}$ is the radius of a primary rolling sphere (solvent molecule), and $R_1$ is the van der Waals radii of atoms.

The procedure of primary rolling may sometimes result in the undesired self-overlapping and kinks of a surface, so the procedure of secondary rolling is applied to smooth them. These irregularities may be classified into the two types:

(1) The self-overlapping of a primary rolling torus; and

(2) The overlapping of concave secondary rolling fragments (a kink).

The diameter of a secondary rolling probe ball is selected so that it is lower than the diameter of a primary rolling probe ball, but exceeds a certain critical distance that depends on the average size of a discrete surface element.

The procedure for the smoothing of kinks and the removal of self-overlapping using the methods of secondary rolling is illustrated in Fig. 2.

The stage of rolling allows us to obtain the formal description of a molecular surface as the sets of the position and orientation coordinates of spherical and toroidal elements and their geometrical connectedness with each other.

The next step in constructing a surface is the generation of a triangulation grid via the sequential addition of triangles. An advantage of this method consists in its universality. It is suitable for any type of surfaces, not merely for surfaces consisting of spherical and toroidal segments. In this method, the algorithm of projecting an arbitrary space point onto the nearest surface point is used. Polygonal surface elements are formed on the basis of triangular surface elements. The centers of polygonal elements coincide with the apices of triangular elements. The apices of polygonal elements are the midpoints of triangular element edges, which originate from the centers of polygonal elements, and the centers of inertia (intersection point of medians) of these triangular elements.

SAS is obtained from SES via similarity transformation: each SAS triangular element is obtained by shifting the three apices of a corresponding triangle on SES by the value of the radius of a primary rolling sphere along the normals to SES at these apices. In this case, only the SAS triangles, whose images on SES have at least one apex on a spherical segment, will have a non-zero surface area.

**4. Algorithms of Solutions**

1. The PCM equation for the charges of surface elements (Eq. (3)) was solved using the iteration scheme [21]: $q_k^{(0)} = \frac{1}{a_{kk}} \sum_i b_{ki} Q_i$, $q_k^{(n+1)} = \frac{1}{a_{kk}} \left( \sum_i b_{ki} Q_i - \sum_{j \neq k} a_{kj} q_j^{(n)} \right)$. The fluctuations of iterations were dumped using the weighted sum of current and previous iterations and the empirical coefficient $q_k^{(n+1)} = (1-\lambda) q_k^{(n+1)} + \lambda q_k^{(n)}$, where $\lambda = 0.35$. To suppress numerical errors, we use Eqs. (5) and (7) that lead to the fulfillment of condition (8) for the sum of surface charge.

2. In a similar fashion, we solve the PCM equation for enlarged surface elements.

3. The COSMO equation for the charges of surface elements was solved by the iteration scheme of the conjugate gradient method that is used for linear equations defined by a symmetric positive-definite matrix [23]. The conjugate gradient method [27] minimizes the energy described by the quadratic equation

$$\Delta G_{pol}(q) = \frac{1}{2} q^T A^c q - f^T q, \tag{19}$$

where $f = -D^T Q$ is the column of the right-hand side of the COSMO equation, and $q$ is the column of the charges of surface elements.

The minimum of energy (19) corresponds to the solution $q$ of the COSMO equation.

4. The polarization energy in the SGB approximation is found directly from Eq. (15) by calculating the Born radii of all the substrate atoms via Eqs. (16) and (17).

**5. Analytical Gradients**

The possibility of using the analytical gradients of the energy of a molecule in a solvent plays an important part in accelerating the optimization of the geometry of a molecule. The present part is devoted to the calculation of gradients. It is obvious that the precision in calculating the gradients must provide a rather good agreement between the analytical and numerical gradients at a relatively small grid step used to calculate the numerical gradients.

For convenience, let us introduce the conjugate charges $q^*$ that satisfy the following equation in both the PCM and COSMO methods:

$$A^T q^* = D^T Q, \tag{20}$$

where the elements of the matrix $D = \{d_{ij}\}$ are determined by Eq. (10).

Multiplying the left-hand side of Eq. (3) by the matrix $A^{-1}$, we find

$$q = A^{-1}BQ. \tag{21}$$

Similarly, from Eq. (20) we have

$$q^* = \left(A^T\right)^{-1} D^T Q. \tag{22}$$

Substituting Eq. (21) for $q$ into Eq. (9), for the PCM method we obtain

$$\Delta G_{pol} = \frac{1}{2} Q^T D A^{-1} B Q. \tag{23}$$

According to Eq. (14), the right-hand side of Eq. (23) in the COSMO method should be multiplied by the factor $C_f$ determined by Eq. (11).

Let us denote the coordinate of the $m^{\text{th}}$ atom as $x_m^k$, where the index $k$ runs over the values 1, 2, 3 and denotes one of the Cartesian coordinates of an atom, and the index $m$ runs over all the numbers of atoms in a molecule. Differentiating both sides of Eq. (23) with respect to the coordinate of an atom $x_m^k$ and using Eqs. (21) and (22), we obtain the expression for the analytical gradients in the PCM method through the polarization $q$ and conjugate $q^*$ charges and the gradients of the matrices $D$, $A$, and $B$:

$$\frac{\partial \Delta G_{pol}}{\partial x_m^k} = \frac{1}{2} Q^T \frac{\partial D}{\partial x_m^k} q - \frac{1}{2} (q^*)^T \frac{\partial A}{\partial x_m^k} q + \frac{1}{2} (q^*)^T \frac{\partial B}{\partial x_m^k} Q. \tag{24}$$

In the COSMO method, the right-hand side of Eq. (24) is multiplied by the factor $C_f$, and the expression itself can be written in a much more compact form with consideration for that, in this method, $B = -D^T$, $q^* = -q$, and $A = A^c$, namely,

$$\frac{\partial \Delta G_{pol}}{\partial x_m^k} = C_f \left( Q^T \frac{\partial D}{\partial x_m^k} q + \frac{1}{2} q^T \frac{\partial A^c}{\partial x_m^k} q \right).$$

The derivatives of the matrices $D$, $A$, and $B$ in the PCM method are found from Eqs. (4), (6), and (10), and the derivatives of the matrices $D$ and $A$ in the COSMO method are determined from Eqs. (10) and (13). Note that, if the atoms, whose positions are used to calculate the gradient, are located near the surface of a solvent, the vectors of surface elements $r_i$ corresponding to the normal and area of surface elements are in general changed, and this must also be taking into consideration, when calculating the gradients.

## 6. Analytical Gradients of the Parameters of Surface Elements

All the surface elements may be classified into the three principal groups: spherical, toroidal, and combined elements. Spherical elements are elements that entirely lie on a single sphere, toroidal elements are elements that entirely lie on a single torus, and combined elements are boundary elements that lie simultaneously on two or more surface fragments. Correspondingly, for each group, there exist peculiar rules of calculating the derivatives. Each surface element is fully described by the following variables: the radius vector of the $i^{th}$ surface element $\vec{r}_i$, the vector of the normal of the $i^{th}$ surface element $\vec{n}_i$, and the area of the $i^{th}$ surface element $S_i$.

A surface is transformed only when the coordinates of surface atoms are changed. The change in the coordinates of a single surface atom entails the transformation of only a small surface domain in the immediate vicinity of this atom. For compactness, the relationships between the total differentials of the parameters of surface elements and the total differentials of the coordinates of atoms are given below. Using such a compact form, we can also obtain the expressions for partial derivatives. Total differentials are denoted by $\Delta$, for example, $\Delta\vec{r}_i$ means the total differential of the radius vector of the $i^{th}$ surface element. The other notations will be explained below in the course of our consideration.

### 6.1. Spherical Elements

The change of the coordinates of spherical elements is completely determined only by the displacement of the center of a sphere, and the changes of the normal and area of a surface element are zero:

$$\Delta\vec{r}_i = \Delta\vec{r}_0, \vec{n}_i = 0, S_i = 0, \tag{25}$$

where $\Delta r_0$ is the differential of the coordinates of the center of a sphere. If an element belongs to the surface of an atomic sphere, the displacement of the center of this sphere is determined by the displacement of the coordinates of an atom, i.e., $\Delta r_0 = \Delta r_1$. In the other cases, there exist more complicated dependences that will be explained below.

### 6.2. Toroidal Elements

For each toroidal surface fragment, we introduce the local orthonormalized basis $x$, $y$, $z$, whose center is located at the point $p_c$ (Fig. 3). The $z$-axis is oriented along the line connecting the centers of atoms: $\vec{z} = \frac{1}{c}(\vec{r}_2 - \vec{r}_1)$, $(\vec{x} \cdot \vec{z}) = 0$, $(\vec{y} \cdot \vec{z}) = 0$, and $(\vec{x} \cdot \vec{y}) = 0$. The apex $p_0$ is found using this local basis as $\vec{p}_0 = \vec{p}_c + h\vec{e}_s$, $\vec{e}_s = \alpha_s \vec{x} + \beta_s \vec{y}$, $\alpha_s^2 + \beta_s^2 = 1$, where $\alpha_s$ and $\beta_s$ are constants that are determined by the normal at the surface point $r_s$:

$\gamma_s = \vec{n}_s \cdot \vec{z}$, $\alpha_s = \dfrac{(\vec{n}_s \cdot \vec{x})}{\sqrt{1-\gamma_s^2}}$, and $\beta_s = \dfrac{(\vec{n}_s \cdot \vec{y})}{\sqrt{1-\gamma_s^2}}$. In Fig. 3, $R_{pr}$ is the radius of a primary rolling sphere, $R_1$ is the radius of the first atom, $R_2$ is the radius of the second atom, and the plane of the figure is specified by the centers of atoms $r_1$ and $r_2$ and a surface point $r_s$. The expressions for the parameters used in Fig. 3 are $a = R_1 + R_{pr}$, $b = R_2 + R_{pr}$, $c = |\vec{r}_2 - \vec{r}_1|$,

$h = \dfrac{1}{2c}\sqrt{4a^2c^2 - (a^2 + c^2 - b^2)^2}$, $\vec{p}_c = \dfrac{1}{2}(\vec{r}_1 + \vec{r}_2) + \dfrac{(\vec{r}_2 - \vec{r}_1)(a^2 - b^2)}{2c^2}$, and $\vec{r}_s = \vec{p}_0 - R_{pr}\vec{n}_s$.

When the coordinates of atoms are changed, the position and orientation of the local basis are also changed, but $\alpha_s$ and $\beta_s$ remain constant.

The differentials of the auxiliary parameters are related with the differentials of the coordinates of atoms $\Delta r_1$ and $\Delta r_2$ as follows:

$$\begin{cases} \Delta c = \dfrac{(\Delta \vec{r}_2 - \Delta \vec{r}_1)(\vec{r}_2 - \vec{r}_1)}{c}, \Delta h = \left(\dfrac{b^2 + a^2 - c^2}{2h} - h\right)\dfrac{\Delta c}{c}, \\ \Delta \vec{p}_c = \dfrac{1}{2}(\Delta \vec{r}_1 + \Delta \vec{r}_2) + \dfrac{(\Delta \vec{r}_2 - \Delta \vec{r}_1)(a^2 - b^2)}{2c^2} - (\vec{r}_2 - \vec{r}_1)(a^2 - b^2)\dfrac{\Delta c}{c^3}, \\ \Delta \vec{z} = \dfrac{(\Delta \vec{r}_2 - \Delta \vec{r}_1)}{c} - \dfrac{(\vec{r}_2 - \vec{r}_1)}{c^2}\Delta c, \Delta \vec{e}_s = -(\vec{e}_s \cdot \Delta \vec{z})\vec{z}. \end{cases}$$

The differentials of the parameters of surface elements are found as

$$\Delta \vec{r}_s = \Delta \vec{p}_s + \Delta h \vec{e}_s + h \Delta \vec{e}_s - R_{pr}\Delta \vec{n}_s, \tag{26}$$

$$\Delta \vec{n}_s = \sqrt{1-\gamma_s^2}\Delta \vec{e}_s + \gamma_s \Delta \vec{z}, \tag{27}$$

$$\Delta S_s = \dfrac{\Delta h}{h - R_{pr}\sqrt{1-\gamma_s^2}}\delta S_s. \tag{28}$$

**6.3. Combined Elements**

Combined elements incorporate the parts of different surface fragments simultaneously. The radius vector of such an element is the radius vector of a certain point that lies on either a spherical or toroidal surface. According to this definition, we formulate the rules of finding the derivative of the coordinates and normal of such a surface element. A different situation arises with the surface area of this element. Each element is composed of several adjacent flat triangles that have a shared apex representing the center of an element. The surface area of a combined element is calculated as the sum of the surface areas of these triangles. Correspondingly, the differential of the surface area of an element is calculated as the sum of the differentials of the surface areas of triangles. A triangle is specified by the apices $\vec{r}_1, \vec{r}_2$, and $\vec{r}_3$, and the generatrix vectors of a triangle are $\vec{a} = \vec{r}_2 - \vec{r}_1$ and $\vec{b} = \vec{r}_3 - \vec{r}_1$. The surface area of a triangle is found as $\vec{s} = [\vec{a} \times \vec{b}]$, $S = \frac{1}{2}|\vec{s}|$, and its differential at $\Delta \vec{a} = \Delta \vec{r}_2 - \Delta \vec{r}_1$ and $\Delta \vec{b} = \Delta \vec{r}_3 - \Delta \vec{r}_1$ has the form

$$\Delta S = \frac{1}{4} \frac{(\Delta \vec{a} \cdot \vec{a})b^2 + (\Delta \vec{b} \cdot \vec{b})a^2 - (\Delta \vec{a} \cdot \vec{b})(\vec{a} \cdot \vec{b}) - (\vec{a} \cdot \vec{b})(\vec{a} \cdot \Delta \vec{b})}{S}. \qquad (29)$$

The differentials of the coordinates of triangle apices are governed by the type of a fragment, on which an apex lies. A molecular surface as a whole consists only of the two types of fragments—spherical and toroidal. The rules for differentiating the parameters of surface elements of each type have been considered above. However, each of these types is divided into several subtypes. Spherical fragments may be formed by both van der Waals atomic spheres and primary or secondary rolling spheres. For each subtype, there exists a peculiar relationship between the coordinates of the center of a sphere and the coordinates of atoms. A similar situation takes place for toroidal elements, which may be formed by both primary and secondary rolling. For the differentials of the elements of secondary rolling toroidal fragments, it is sufficient to apply above derived Eqs. (26) – (29), replacing the coordinates of atoms by the coordinates of the centers of corresponding primary rolling spheres and switching a normal to the opposite direction.

Spherical fragments may be represented by fragments of both primary and secondary rolling. In the case of primary rolling, such a fragment is a spherical triangle formed by a rolling sphere supported by three atoms. If $\vec{r}_1$, $\vec{r}_2$, and $\vec{r}_3$ are the coordinates of atoms and $\vec{r}_0$ is the coordinates of the center of a sphere, the relationship between the differentials has the form

$$\Delta \vec{r}_0 = \frac{\left(\Delta \vec{r}_1 \cdot (\vec{r}_1 - \vec{r}_0)\right)}{\left(\left[(\vec{r}_2 - \vec{r}_0) \times (\vec{r}_3 - \vec{r}_0)\right] \cdot (\vec{r}_1 - \vec{r}_0)\right)} \left[(\vec{r}_2 - \vec{r}_0) \times (\vec{r}_3 - \vec{r}_0)\right] + $$
$$+ \frac{\left(\Delta \vec{r}_2 \cdot (\vec{r}_2 - \vec{r}_0)\right)}{\left(\left[(\vec{r}_1 - \vec{r}_0) \times (\vec{r}_3 - \vec{r}_0)\right] \cdot (\vec{r}_2 - \vec{r}_0)\right)} \left[(\vec{r}_1 - \vec{r}_0) \times (\vec{r}_3 - \vec{r}_0)\right] + \quad (30)$$
$$+ \frac{\left(\Delta \vec{r}_3 (\vec{r}_3 - \vec{r}_0)\right)}{\left(\left[(\vec{r}_1 - \vec{r}_0) \times (\vec{r}_2 - \vec{r}_0)\right] (\vec{r}_3 - \vec{r}_0)\right)} \left[(\vec{r}_1 - \vec{r}_0) \times (\vec{r}_2 - \vec{r}_0)\right].$$

In the case of secondary rolling, there exist spherical fragments of the two types. The first type is similar to spherical fragments of primary rolling, and Eq. (30) is applicable to the coordinates of its center, but the coordinates of atoms will be replaced by the corresponding coordinates of the centers of primary rolling spheres. The second type corresponds to the positions of a secondary rolling sphere supported by the converging neck of a primary rolling torus (Fig. 4). The coordinates of the center of a sphere are $\vec{r}_0 = \vec{p}_c \pm d\vec{z}$ and $d = \sqrt{(R_{pr} + R_{\sec})^2 - h^2}$, where $R_{\sec}$ is the radius of a secondary rolling sphere. The differentials have the form

$$\Delta d = -\frac{h \Delta h}{d}, \quad \Delta \vec{r}_0 = \Delta \vec{p}_c \pm (\Delta d \vec{z} + d \Delta \vec{z}), \quad (31)$$

where we used the notations that are precisely the same as in the case of the above described differentials of toroidal elements.

**7. Analytical Gradients of the Energy in the SGB Model**

The total energy differential at constant charges and dielectric permittivities is found via the direct differentiation of Eq. (15) for the energy in the SGB model as

$$\Delta(\Delta G_{pol}) = -\frac{1}{2}\left(1 - \frac{1}{\varepsilon}\right) \sum_{i,j} Q_i Q_j \Delta g_{i,j}, \quad (32)$$

where the functions $g_{i,j}$ are determined as $g_{i,j} = \dfrac{1}{\sqrt{\left|\vec{R}_{i,j}\right|^2 + a_i a_j \exp\left(-\dfrac{\left|\vec{R}_{i,j}\right|^2}{c a_i a_j}\right)}}$, and $\vec{R}_{i,j}$ is the

vector between the $i^{th}$ and $j^{th}$ atoms. The total differential has the form

$$\Delta g_{i,j} = \dfrac{-g_{i,j}^3}{2}\left[2\left(\vec{R}_{i,j}\cdot\Delta\vec{R}_{i,j}\right)\left(1 - \dfrac{1}{c}\exp\left(-\dfrac{\left|\vec{R}_{i,j}\right|^2}{c a_i a_j}\right)\right) + \left(\Delta a_i a_j + a_i \Delta a_j\right)\left(1 + \dfrac{\left|\vec{R}_{i,j}\right|^2}{c a_i a_j}\right)\exp\left(-\dfrac{\left|\vec{R}_{i,j}\right|^2}{c a_i a_j}\right)\right]. \quad (33)$$

The differentials for the Born radius of the $i^{th}$ atom are determined as

$$\Delta a_i = -\dfrac{1}{2}\left(A_0 + \sum_n A_n I_n^i\right)^{-2}\sum_n A_n \Delta I_n^i, \quad 7 \geq n \geq 4. \quad (34)$$

The differentials of the surface integrals are found as

$$\Delta I_n^i = \dfrac{\Delta J_n^i}{n-3}\left[J_n^i\right]^{(4-n)/(n-3)}, \quad 7 \geq n \geq 4,$$

where $\Delta J_n^i = \sum_s \Delta J_{s,n}^i$, $J_{s,n}^i = \dfrac{\left(\vec{n}_s \cdot (\vec{r}_s - \vec{r}_i)\right) S_s}{\left|\vec{r}_s - \vec{r}_i\right|^n}$, $s$ is the number of a surface element, $i$ is the

number of an atom, and

$$\Delta J_{s,n}^i = \dfrac{\left[\left(\vec{n}_s \cdot (\Delta\vec{r}_s - \Delta\vec{r}_i)\right) + \left(\Delta\vec{n}_s \cdot (\Delta\vec{r}_s - \Delta\vec{r}_i)\right)\right] S_s + \left(\vec{n}_s \cdot (\vec{r}_s - \vec{r}_i)\right)\Delta S_s}{\left|\vec{r}_s - \vec{r}_i\right|^n} -$$
$$- \dfrac{n\left(\vec{n}_s (\vec{r}_s - \vec{r}_i)\right)\left((\Delta\vec{r}_s - \Delta\vec{r}_i)\cdot(\vec{r}_s - \vec{r}_i)\right) S_s}{\left|\vec{r}_s - \vec{r}_i\right|^{n+2}}. \quad (35)$$

The derivatives are calculated by Eqs. (32)–(35) on the basis of the surface element derivatives that are used in Eq. (35) and previously found by Eqs. (25)–(31).

### 8. Gradients of the Non-Polar Component of the Solvation Energy

The gradient of the non-polar component of the solvation energy is expressed as

$$\Delta(\Delta G_{nonpol}) = \sigma \sum_{j=1}^{N} \Delta S_j,$$

where the index $j = 1, \ldots, N$ runs over all the SAS elements. A non-zero contribution to the values of $\Delta S_j$ is made only by the surface elements that lie near the boundary of spherical fragments. The gradients of these elements are determined by Eqs. (25) and (29), as SAS consists of spherical fragments alone.

### 9. Software Validation

All the methods implemented in the DISOLV software require the construction of SES, its triangulation, and the solution of the corresponding equations on this surface. The model parameters are the step of a triangulation grid and the positions of its nodes, the primary rolling radius, and also the parameters used to smooth a surface, such as the secondary rolling radius and the critical distance. First of all, validation must show that we can select a triangulation grid step such that its arbitrary shift along SES produces a rather small change in the polarization energy within the precision of 0.1–1 kcal/mol that is required for the calculation of the protein–ligand binding energy.

Validation was performed at a fixed primary rolling sphere radius $R_{pr}$ = 1.4 Å [17, 28] and a solvent (room temperature water) dielectric permittivity $\varepsilon$ = 78.5. When selecting the values of the other model parameters, we assume that

(1) For a triangulated surface to be rather smooth, the triangulation grid step must be smaller than the radii of atoms, on which a surface is constructed;

(2) The radius of a secondary rolling sphere must exceed the triangulation grid step (to reproduce the smooth features of a surface during its triangulation), but smaller than the primary rolling diameter and the diameter of atoms (to prevent any considerable deformation of a surface due to secondary rolling); and

(3) The critical distance must be selected such that it would be much smaller than the radius of a secondary rolling sphere, since the critical distance is a certain minimally admissible surface curvature criterion, upon the satisfaction of which a surface is smoothed to the value of secondary rolling.

The secondary rolling radius and the critical distance were selected as equal to 0.4 and 0.15 Å, respectively. Even relatively wide variations of these parameters did not appreciably worsen the precision of calculations (< 0.2 kcal/mol).

The root-mean-square deviation (RMSD) from the average value of the polar component of the protein–ligand complex desolvation energy calculated by different methods as a function of the triangulation grid step is plotted in Fig. 5. Variations in the desolvation energy occurred, when a grid was shifted along SES as specified by the values of the two angles that determined the initial grid generation point and was evenly changed, taking 200 values each within their definition ranges. The averaging over these variations gave the average value and RMSD of the desolvation energy. For our calculations, we used the protein–ligand complex taken from the Protein Data Bank (PDB) [29], PDB entry 1SQO, that contained a native ligand of 34 atoms with a total charge of +1 and a protein (urokinase) of 3818 atoms with a total charge of +3. Hydrogen atoms absent in PDB were added to the protein with the Reduce software [30] and to the ligand with the MolRed software [20, 31], and atomic charges were positioned in accordance with the MMFF94 force field [15] using the SOL docking software [2].

From Fig. 5 it can be seen that the root-mean-square deviation RMSD from the average desolvation energy monotonically tends to zero with decreasing grid step for all the four methods of calculation. The average desolvation energy itself arrives at a constant value for each of the four methods (omitted in Fig. 5). For the PCM, COSMO, and SGB methods, the error introduced by variations in the position of a triangulation grid on SES becomes lower than 0.5 kcal/mol even at a grid step of 0.3 Å and lower than 0.2 kcal/mol at a grid step below 0.2 Å. For the Abagyan method of enlarged elements, this error is appreciably higher and amounts 0.5 kcal/mol even at a grid step of 0.15 Å.

Further we compared the desolvation energies for the methods PCM, SGB, COSMO, and PCM with enlarged elements and 20 complexes taken from PDB at a grid step of 0.2 Å, at which all these methods gave an error of less than 1 kcal/mol for the 1SQO complex (Fig. 5). The complexes were selected such that they were rather diverse and contained from 3000 to 10000 atoms with a scatter in the charges of proteins and ligands from −15 to +8 and from −2 to +2, respectively. The corresponding structures were prepared as described above for urokinase. The mean-square deviation from the desolvation energy calculated by the most precise PCM method was 4.5, 5.7, and 6.3% for COSMO, SGB, and PCM with enlarged elements, respectively.

The correctness of the calculation of analytical gradients was checked by means of their comparison with numerical gradients. This was performed for thrombase, whose structure was taken from PDB (1O2G) [29] and prepared as described above for urokinase. At a grid step of 0.1 Å, the difference between the numerical and analytical gradients for the PCM, COSMO, and SGB methods was found to be equal to several percents for superficial atoms and two orders of magnitude smaller for volumetric atoms.

In the comparison of the calculated and experimentally measured solvation energies of molecules, an important role belongs to the selection of atomic radii that specify SES. As a basis, we took the corresponding radii from [25]. However, in contrast to [25], we refused to use one more parameter—the intramolecular dielectric permittivity, which is distinct from unity and taken in [25] as $\varepsilon_{in} = 2.21$. We used $\varepsilon_{in} = 1$ at a water dielectric permittivity $\varepsilon = 78.5$ and corrected the radii from [25] so as to minimize the root-mean-square deviation of calculated solvation energies, including both the polar (PCM) and non-polar components, from experimental values for 410 molecules considered in [25]. The coefficients $a$ and $b$ incorporated into Eq. (18) for the non-polar solvation energy component were also determined from the condition of a minimum of the above mentioned root-mean-square deviation. The calculated and experimental solvation energies are compared in Fig. 6, and the found optimal atomic radii are represented in Table 1 together with the radii from [25]. The optimal coefficients $a$ and $b$ are equal to 0.00387 kcal/(mol Å$^2$) and 0.698 kcal/mol, respectively.

From Table 1 it follows that the passage from the intramolecular dielectric permittivity $\varepsilon_{in} = 2.21$ (as in [25]) to $\varepsilon_{in} = 1$ (this work) leads to a certain increase of atomic radii, as the polarization charges on SES grow with a decrease in the intramolecular dielectric permittivity and, to keep the polarization component of interaction between molecular and polarization charges at the same level, the position of SES must be slightly shifted from a molecule, and this means that the atomic radii must be increased.

Unfortunately, not all the MMFF94 types of atoms are represented in the set of molecules from [25]. For example, the types inherent to molecular ions are absolutely absent in this set. The atomic radii for these types of atoms were obtained by minimizing the root-mean-square deviation of calculated solvation energies from their experimental values taken from [33, 34], assuming the earlier found atomic radii to be constant. Thus calculated solvation energies and their experimental values are compared in Fig. 7, and the additionally obtained radii of atoms and their types are listed in Table 2. The root-mean-square deviation between calculated and experimental values is equal to 9.81 kcal/mol, and this disagreement between theory and experiment lies within the limits of experimental errors.

Finally, we compared the results of calculations by the DISOLV software with the analytically solved case of a point charge inside a spherical cavity in a dielectric. The polarization energy determined as the difference between the electrostatic energy of a point charge in a spherical cavity of a dielectric and the electrostatic energy of the same charge in a vacuum has the form

$$E = -\sum_{n=0}^{\infty} (\varepsilon - 1) \frac{Q^2 (n+1) r^{2n}}{2(n\varepsilon + \varepsilon + n) a^{2n+1}}, \qquad (36)$$

where $\varepsilon$ is the dielectric permittivity of a medium and equal to unity inside a cavity, $a$ is the radius of a spherical cavity, and $r$ is the distance from the center of a sphere to the point charge $Q$. The gradient along the normal to a surface can easily be expressed by differentiating Eq. (36) as

$$\frac{\partial E}{\partial r} = -\sum_{n=1}^{\infty} (\varepsilon - 1) \frac{Q^2 n (n+1) r^{2n-1}}{(n\varepsilon + \varepsilon + n) a^{2n+1}}. \qquad (37)$$

In the DISOLV software, such a configuration was modeled with two fictitious atoms, one of which was neutral and had a radius of 6 Å, and the other had a unity charge and a radius of 0.5 Å. The results of calculations by Eqs. (36) and (37) and by the DISOLV software at a grid step of 0.1 Å for different positions of a unit charge are given in Table 3.

As can be seen, the results of calculation by Eqs. (36) and (37) are reproduced by the DISOLV software with a high precision in the PCM method, and an exception is the gradient for a charge that is located at a distance of 1 Å from the boundary of a dielectric, when the selected grid step is rather small, but, in this case, the difference between the precise gradient and its value calculated by the DISOLV software also amounts only several percents.

The time $t$ of calculating the solvation energy for proteins of several thousands of atoms without gradients depends on the grid step $h$ as $t \approx h^{-4}$ for the methods PCM and COSMO and as $t \approx h^{-2.5}$ and $t \approx h^{-3}$ for the methods SGB and PCM with enlarged elements, respectively. For example, PCM, the most precise among the used methods, requires nearly 600 s per 1 CPU (Intel Xeon E5472, 3.0 GHz) to calculate the solvation energy of urokinase incorporated into the PDB 1SQO complex.

The comparison of the results of calculating the solvation energy of a protein by the DISOLV software using the PCM method and by the APBS software [14] at the same atomic radii and charges, grid step, and precision of calculations shows that the difference between the energies is several percents and, with respect to time, the DISOLV software is an order of magnitude faster.

**10. Conclusions**

In the present work, we considered one of the aspects of improving the precision of the computer-aided prediction of the inhibiting activity of molecules that are candidates to inhibitors for specified target proteins, namely, the allowance for the effect of a solvent on the protein–ligand binding energy. The physical foundations of the four implicit (continual) solvent models implemented in the DISOLV software, such as PCM, COSMO, SGB, and PCM with enlarged elements, were described in details, and the methods of solving the corresponding equations and the foundations of the algorithm used to construct the surfaces in these models and calculate the gradients of the energy of a molecule in a solvent were discussed. The latter gradients are needed in the DISOLV software for the local optimization of the energy of a molecule in a solution. The DISOLV validation results that showed not only the possibility of attaining a good precision of calculations (at arbitrary shifts of a triangulation grid) of better that several tenths of kilocalories per mole over reasonable time periods for such large macromolecules as proteins, but also a good agreement (root-mean-square deviation, 0.8 kcal/mol) of the calculated Gibbs energies of dissolution of a molecule in water, i.e., the energy of transfer of a molecule from a vacuum into water, with their experimental values for several hundreds of neutral molecules were represented. For molecular ions, the root-mean-square deviation between the calculated and experimental solvation energies is considerably higher (10 kcal/mol), but this value also lies within the limits of measurement error in most cases. On the whole, the validation results show that the DISOLV software can be used in the post-processing regime to refine the protein–ligand binding energy estimate given by the docking program.

Table 1. MMFF94 force field atomic types and the atomic radii corresponding to them from [25] and the atomic radii found in this work

| MMFF94 atomic types [15] | Atomic radii from [25], Å | Atomic radii obtained in this work, Å | MMFF94 atomic types [15] | Atomic radii from [25], Å | Atomic radii obtained in this work, Å |
|---|---|---|---|---|---|
| C_1 | 1.46 | 1.95 | C_2 | 1.92 | 2.33 |
| C_3 | 2.18 | 2.77 | C_4 | 1.84 | 2.02 |
| H_5 | 1.18 | 1.14 | O_6 | 1.38 | 1.66 |
| O_7 | 1.35 | 1.49 | N_8 | 1.17 | 1.24 |
| F_11 | 2.14 | 2.79 | Cl_12 | 1.93 | 2.44 |
| Br_13 | 1.55 | 1.85 | I_14 | 1.17 | 1.35 |
| S_15 | 1.62 | 1.96 | H_21 | 0.96 | 1.03 |
| C_22 | 1.91 | 2.18 | H_23 | 1.34 | 1.71 |
| H_24 | 1.40 | 1.56 | H_28 | 0.94 | 1.01 |
| H_29 | 1.01 | 1.07 | O_32 | 1.76 | 2.15 |
| C_37 | 1.71 | 2.19 | N_38 | 1.70 | 1.96 |
| N_40 | 1.50 | 1.68 | N_42 | 1.90 | 2.19 |
| N_45 | 1.87 | 2.16 | | | |

Table 2. Atomic radii for the types of atoms that are encountered in molecular ions and absent in the set of 410 neutral molecules [25]

| MMFF94 atomic type | Radius, Å |
|---|---|
| N_34 | 2.04 |
| O_35 | 1.83 |
| H_36 | 1.25 |
| C_41 | 1.98 |
| O_49 | 2.14 |
| H_50 | 0.88 |
| O_51 | 1.71 |
| H_52 | 0.65 |

Table 3. Polarization energy and its gradients calculated by Eqs. (36) and (37) and by the DISOLV software for a unit point charge inside a spherical dielectric cavity of 6 Å in radius with a dielectric permittivity $\varepsilon = 78.5$ at different positions of this charge, $r$ is the distance from a charge to the center of a spherical cavity

| $r$, Å | Energy, kcal/mol | | Gradient, kcal/(mol Å) | | |
|---|---|---|---|---|---|
| | Eq. (36) | DISOLV | Eq. (37) | DISOLV, analytical | DISOLV, numerical |
| 0 | −27.3201 | −27.3201 | 0 | 1.83E−05 | 2.84E−05 |
| 0.2 | −27.3503 | −27.3503 | −0.302306 | −0.30257 | −0.30262 |
| 3 | −36.3637 | −36.3632 | −8.03456 | −8.03357 | −8.03618 |
| 5 | −88.8714 | −88.8452 | −80.4738 | −81.45514 | −74.56199 |

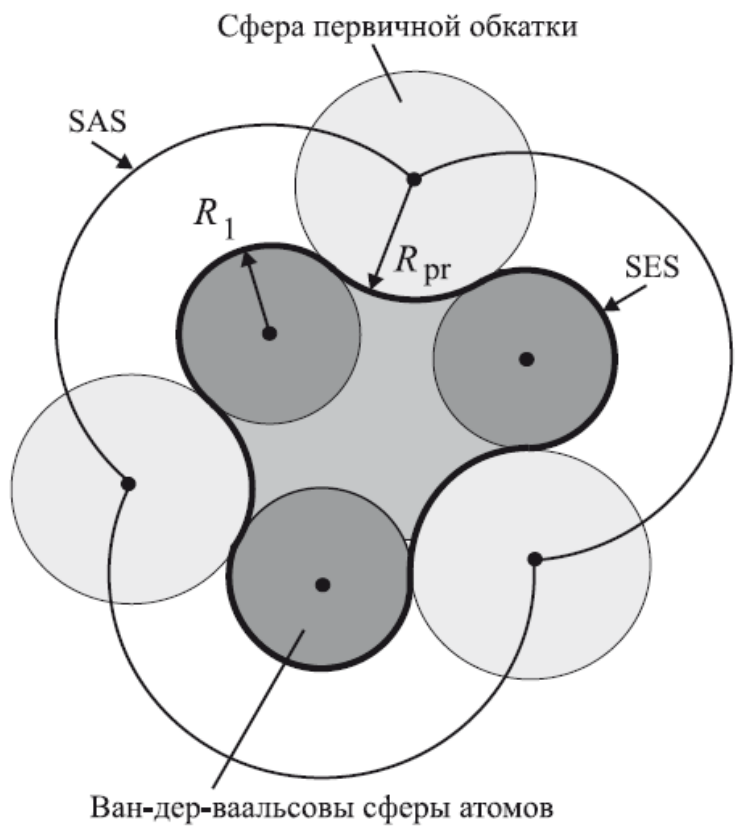

Fig.1. Primary rolling of a surface.

Key:

Сфера первичной обкатки --> Primary rolling sphere;

Ван-дер-Ваальсовы сферы атомов --> van der Waals atomic spheres

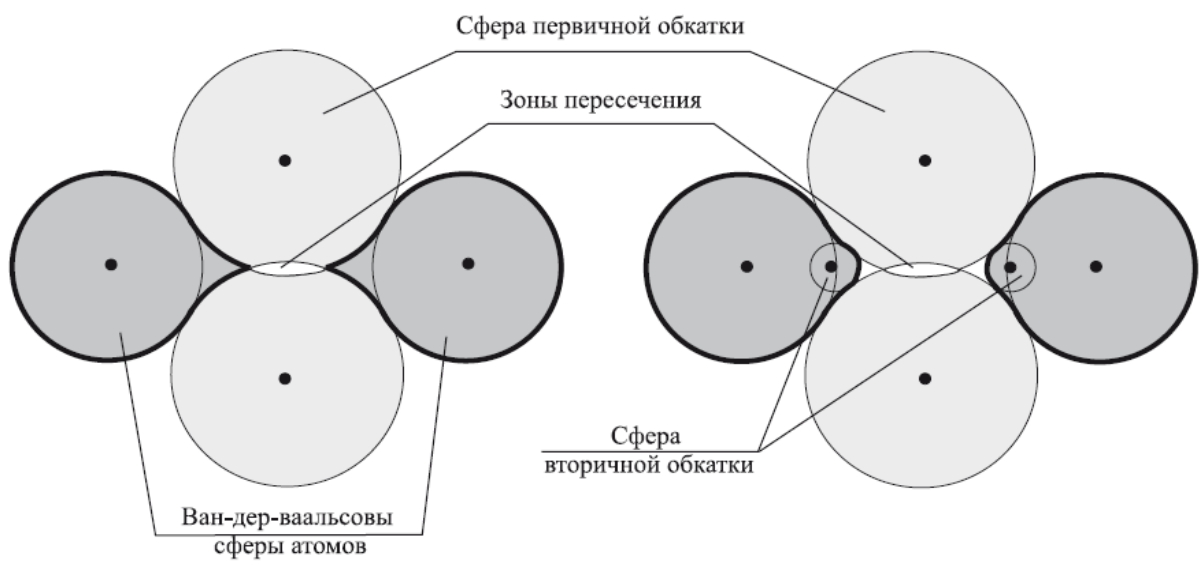

Fig. 2. Secondary rolling of the self-overlapping of a primary rolling torus.

Key:

Сфера первичной обкатки --> Primary rolling sphere;

Зоны пересечения --> Overlapping zones;

Сфера вторичной обкатки --> Secondary rolling sphere;

Ван-дер-Ваальсовы сферы атомов --> Van der Waals atomic spheres

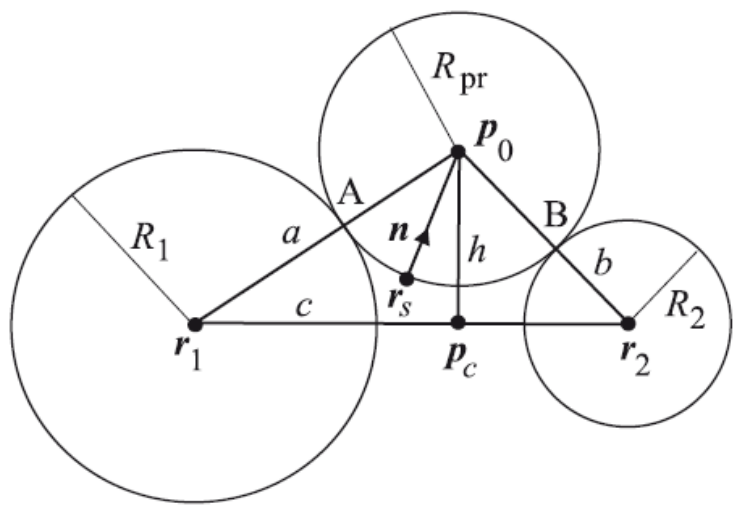

Fig. 3. Geometric configuration of atoms and a primary rolling sphere in the formation of a toroidal surface fragment

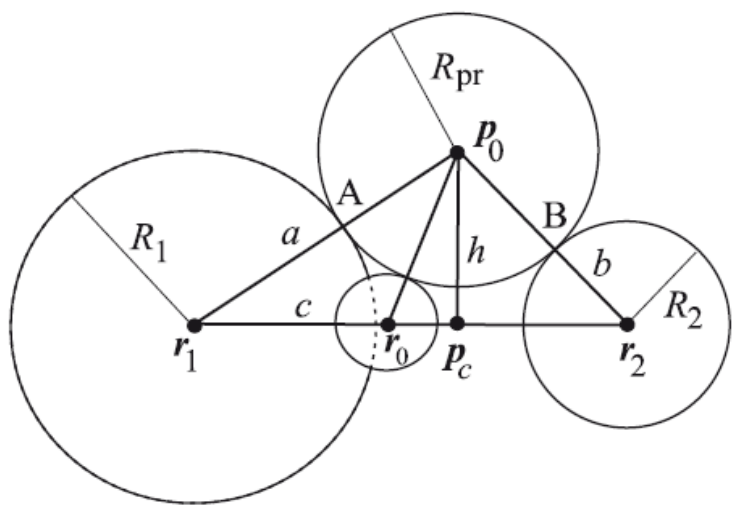

Fig. 4. Schematic representation of the geometric configuration, in which a secondary rolling sphere $r_0$ is supported by the converging neck AB of a primary rolling torus.

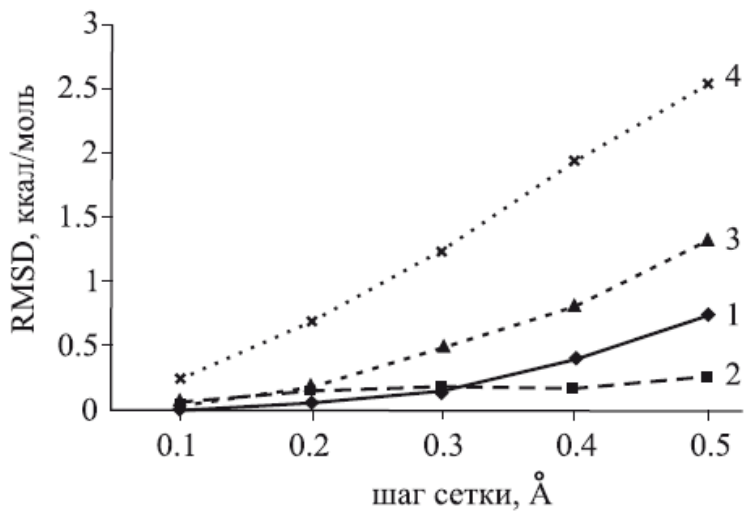

Fig. 5. Dependence of the root-mean-square deviation RMSD from the average desolvation energy of the urokinase–ligand complex (PDB entry 1SQO) (due to the change of grid orientation) on the grid step for the methods (*1*) SGB, (*2*) COSMO, (*3*) PCM, and (*4*) PCM with enlarged elements.

Key:
RMSD, ккал/моль --> RMSD, kcal/mol;
шаг сетки, Å --> grid step, Å

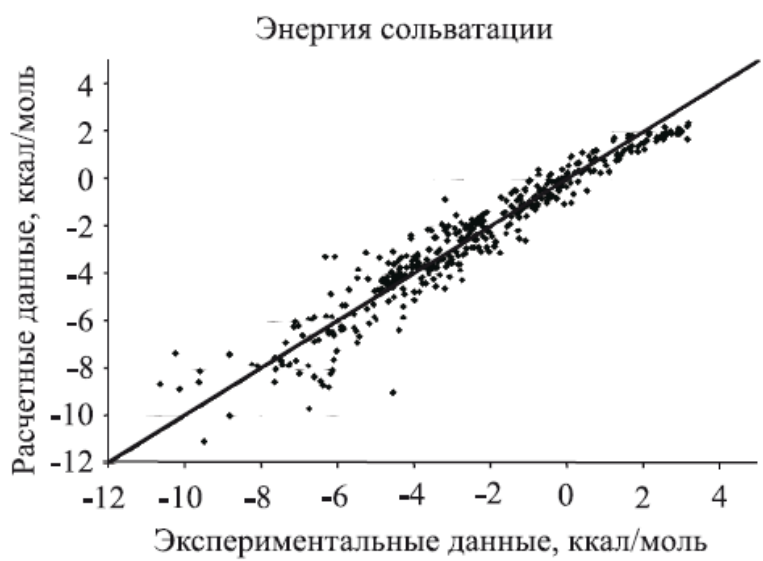

Fig. 6. Comparison of the results of calculating (by PCM for the polar component) the solvation energy (more properly, the difference between the energies of a molecule in a vacuum and water) with experiment for 410 neutral molecules from [25]. The root-mean-square deviation between calculated and experimental values is 0.827 kcal/mol.

Key:

Расчетные данные, ккал/моль --> Calculated data, kcal/mol;

Экспериментальные данные, ккал/моль --> Experimental data, kcal/mol;

Энергия сольватации --> Solvation energy

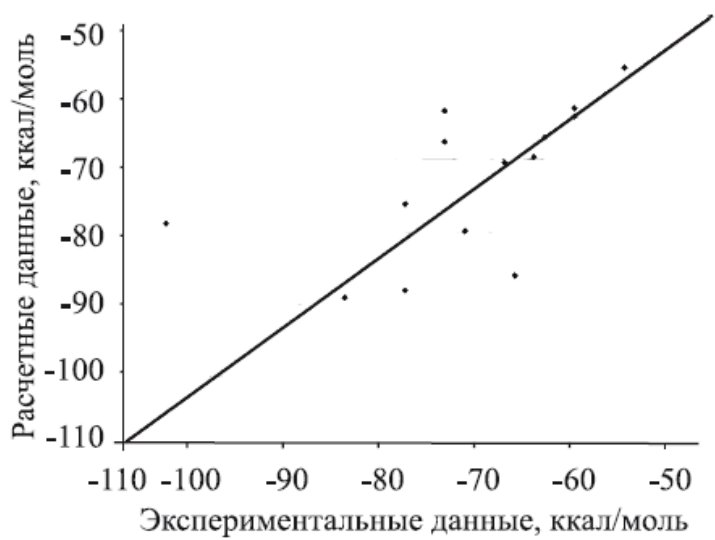

Fig. 7. Comparison of the results of calculating (by PCM for the polar component) the solvation energy with experiment for the molecular ions from [33, 34].

Key:

Расчетные данные, ккал/моль --> Calculated data, kcal/mol;

Экспериментальные данные, ккал/моль --> Experimental data, kcal/mol

# Континуальная модель растворителя: программа *DISOLV* – алгоритмы, реализация и валидация

О.Ю.Купервассер[1], С.Н.Жабин[1], Я.Б.Мартынов[1], К.М.Федулов[1], И.В.Офёркин[1], А.В.Сулимов[1], В.Б.Сулимов[1]

[1] Научно-исследовательский вычислительный центр Московского Государственого Университета имени М.В. Ломоносова, Россия, 119992, Москва, Ленинские горы 1, строение 4; ООО «Димонта», 117186, ул. Нагорная, дом 15, строение 8, Москва, Россия
О.Ю.Купервассер, старший научный сотрудник, e-mail: olegkup@yahoo.com; С.Н.Жабин, младший научный сотрудник, e-mail: delafrog@gmail.com; Я.Б.Мартынов, старший научный сотрудник, e-mail: yaroslavmartynov@yandex.ru; К.М.Федулов, программист, e-mail: cf@dimonta.com; И.В.Офёркин, программист, e-mail: io@dimonta.com; А.В.Сулимов, системный программист, e-mail: sulimovv@mail.ru; В.Б.Сулимов, зав. лабораторией, научный руководитель, e-mail: vladimir.sulimov@gmail.com.

## 1. Введение.

Главная парадигма, используемая при современной разработке новых лекарств, заключается в следующем. Многие болезни связаны с функционированием определенных белков. Поэтому для излечения заболеваний, надо блокировать работу этих белков. Например, белок-мишень может принадлежать вирусу, и его блокирование позволяет сделать невозможным размножение вируса в организме. Блокирование осуществляется с помощью молекул, которые избирательно связываются с этими белками в организме. Такие молекулы, составляющие основу новых лекарств, называются ингибиторами. Как правило, ингибиторы – это сравнительно небольшие органические молекулы, связывающиеся с определенными областями белков-мишеней. Эти области называются центрами связывания или активными центрами. Поиск таких молекул-ингибиторов для заданного белка-мишени и составляет начальный этап разработки нового лекарства. Быстрое и эффективное решение этой задачи в значительной степени определяет минимизацию материальных затрат и продолжительность последующих этапов разработки нового лекарства. По времени этап разработки новых ингибиторов занимает примерно 50% от общей длительности разработки нового лекарства.

Существенно сократить затраты времени и средств на этапе поиска ингибиторов можно с помощью методов компьютерного молекулярного моделирования [1], среди которых главную роль играет докинг. Докинг – это позиционирование молекул кандидатов в ингибиторы (их часто называют лигандами от латинского слова ligare – связываться) в активном центре белка-мишени и оценка их энергии связывания. Чем сильнее молекула связывается с белком, тем лучше ингибитор и эффективнее новое лекарство на его основе. Докинг осуществляется специальными программами молекулярного моделирования (см., например, [2]), которые используются, в том числе и на суперкомпьютерах.

Точность оценки энергии связывания белок-ингибитор определяет эффективность предсказания активности ингибитора: чем больше энергия связывания, тем выше активность ингибитора, и тем эффективнее лекарство на его основе, поскольку требуемого эффекта можно добиться при использовании меньшей концентрации лекарства. Если точность расчетов энергии связывания лиганда с белком недостаточно высока, то вероятность предсказания того, что новые синтезированные соединения – лиганды – ингибируют этот белок, будет низка, и большое количество средств, использованных на синтез лигандов и измерения их ингибирующей активности, и соответствующее время будут потрачены впустую. Достаточно высокая практическая предсказуемость достигается при точности расчета энергии связывания белок-ингибитор

≤ 1 kcal/mol. Поэтому при молекулярном моделировании взаимодействия лигандов с белками-мишенями точность вычисления всех вкладов в энергию связывания белок-лиганд должна быть максимально высокой.

На точность оценки энергии связывания белок-ингибитор при молекулярном моделировании влияют много факторов, например, качество используемого силового поля для описания внутримолекулярных и межмолекулярных взаимодействий, эффективность поиска глобального минимума при выполнении позиционирования ингибитора в активном центре белка-мишени, оценка вклада энтропийной составляющей в свободную энергию связывания белок-ингибитор и др. Поскольку процесс связывания белок-ингибитор в экспериментах (*in vitro* и *in vivo*) имеет место в водном растворе, то при вычислении энергии связывания белок-лиганд надо учитывать и наличие растворителя - воды.

Влияние растворителя на энергию связывания лиганда с белком определяется в основном через энергию десольватации, представляющую собой разность энергии сольватации комплекса белок-лиганд и энергий сольватации белка и лиганда по отдельности. Этот вклад в энергию связывания белок-лиганд обусловлен тем, что при связывании лиганда с белком растворитель (вода) вытесняется из активного центра лиганда, и при этом часть атомов лиганда и часть атомов активного центра белка перестают взаимодействовать с растворителем. Таким образом, для вычисления энергии десольватации необходимо вычислить энергии сольватации белка, лиганда и их комплекса.

Для подсчета свободной энергии сольватации необходимо построить модель растворителя. Это можно сделать явно – путем рассмотрения растворителя как набора большого числа молекул. Однако этот метод при моделировании требует сравнительно больших затрат вычислительных ресурсов, поскольку при вычислении наблюдаемых эффектов необходимо провести усреднение по состоянию молекул растворителя, например с помощью методов молекулярной динамики или Монте Карло. Поэтому более часто используются неявные (их часто называют континуальными) модели растворителя, в которых растворитель рассматривается как непрерывная (континуальная) среда с заданными свойствами, в том числе с заданной диэлектрической проницаемостью [3-5].

Настоящая статья посвящена программе *DISOLV* [6, 7], позволяющей рассчитывать свободную энергию сольватации молекул и градиенты от этой энергии по смещениям атомов молекул, поскольку программа *DISOLV* будет использоваться для оптимизации структуры молекул в растворителе, как с помощью методов силового поля, так и квантовохимических методов, где обычно применяются алгоритмы локальной оптимизации, использующие градиенты.

Свободная энергия Гиббса для процесса сольватации, т.е. перехода молекулы из вакуума в растворитель, или коротко – энергия сольватации, $\Delta G_s$, представляется в виде суммы трех составляющих

$$\Delta G_s = \Delta G_{pol} + \Delta G_{np} + \Delta G_{cav},$$

где $\Delta G_{pol}$ – полярная составляющая взаимодействия молекулы растворяемого вещества (субстрата) с растворителем, $\Delta G_{np}$ – неполярная часть взаимодействия молекулы растворяемого вещества с растворителем, обусловленная Ван-дер-Ваальсовыми силами межмолекулярного взаимодействия, $\Delta G_{cav}$ - кавитационная составляющая свободной энергии растворения, обусловленная образованием в объеме растворителя полости, в которой находится молекула растворенного вещества.

В программе *DISOLV* основное внимание уделено вычислению полярной составляющей взаимодействия молекулы с растворителем $\Delta G_{pol}$ (или коротко энергии поляризации), для вычисления которой используются несколько методов, а остальные составляющие энергии сольватации $\Delta G_{np}$ и $\Delta G_{cav}$ учтены простым и довольно распространенным способом (см. ниже).

В рамках используемой континуальной модели энергия поляризации представляет собой энергию электростатического взаимодействия зарядов атомов молекулы, находящейся в полости диэлектрика, с наведёнными ими на этой поверхности поверхностными зарядами.

К настоящему времени многие существующие неявные модели учета растворителя и их программные реализации интегрированы в более крупные пакеты, например, в квантовохимические пакеты Gaussian [8], Gamess [9], MolPro [10], MOPAC [11], в пакет молекулярной динамики Charmm [12], а имеющиеся в свободном доступе программы для нахождения полярной части взаимодействия с растворителем, например, DelPhi [13] или APBS [14] реализуют численное решение конечно-разностной аппроксимации 3-х мерного уравнения Пуассона-Больцмана.

Целью настоящей статьи является изложение оригинальных алгоритмов вычисления полярной составляющей энергии сольватации молекул, ориентированных на решение соответствующих уравнений на двумерной поверхности окружающего молекулу растворителя, их программной реализации *DISOLV*, написанной на языке C++, и соответствующей валидации. В конце статьи приводится краткое сравнение характеристик программы *DISOLV* с соответствующими характеристиками программы APBS [14], реализующей численное решение уравнения Пуассона-Больцмана в 3D-пространстве.

## 2. Физические модели.

В приближении континуальной модели электростатическое взаимодействие молекулы с растворителем, описываемое членом ($\Delta G_{pol}$), оценивается как взаимодействие точечных зарядов, расположенных внутри полости вырезанной в объёме однородного диэлектрика или проводника (см. ниже), с поляризационными зарядами на поверхности диэлектрика. Роль растворителя играет диэлектрик или проводник. Форма вырезанной в нём полости соответствует форме молекулы, с точностью до размера молекулы растворителя. Заряды в полости описывают неравномерное распределение электронной плотности внутри молекулы. Обычно заряды располагают в центрах атомов, образующих растворяемую молекулу. Заряды находятся либо из квантово-химических расчетов заменой электронного распределенного заряда набором эквивалентных точечных зарядов, либо в рамках модели силового поля, т.е. набора классических потенциалов, описывающих взаимодействие между атомами, когда они определяются соответствующей типизацией атомов. В настоящей работе атомные заряды молекул определяются силовым полем MMFF94 [15].

Однако, чтобы решить электростатическую задачу, необходимо предварительно определить, что такое поверхность молекулы. Существует два определения поверхности, окружающей молекулу [16, 17]: SES (Solvent Excluded Surface) - поверхность исключённого из растворителя объёма. Объем, занимаемый растворителем, лежит *вне* объема, ограниченного этой поверхностью. Сам субстрат полностью лежит *внутри* этого объема. SAS (Solvent Accessible Surface) - поверхность доступная растворителю образуется центрами молекул растворителя, касающихся молекулы субстрата.

Поверхность SES соответствует границе между областями растворяемой молекулы и растворителя, т.е. это контактная поверхность атомов молекулы растворяемого вещества и молекул растворителя. Именно на этой поверхности и должны располагаться индуцированные в диэлектрике заряды, исходя из их физического смысла. Поэтому поверхность SES используется именно для электростатических расчётов.

В отличие от SES поверхность типа SAS описывает границу, вблизи которой средняя концентрация молекул растворителя меняется от нуля до своего среднего по объёму растворителя значения. Поскольку неэлектростатическая составляющая энергии описывается числом молекул растворителя, соприкасающихся с поверхностью субстрата, это число пропорционально площади SAS, и поэтому неэлектростатическая составляющая энергии часто описывается линейной формой от площади SAS [3]. В программе *DISOLV* для построения как поверхности SES, так и поверхности SAS используется модифицированная программа TAGSS [18-20].

В программе *DISOLV* были использованы следующие четыре метода вычисления полярной составляющей взаимодействия молекулы растворяемого вещества с растворителем в континуальном приближении:

1. *PCM* – Polarized Continuum Model [3, 4]. Этот метод представляет собой точное решение электростатической задачи, когда точечные заряды, локализованные в определенных точках пространства – там, где находятся соответствующие атомы молекулы растворяемого вещества, окружены непрерывным диэлектриком с заданной диэлектрической проницаемостью $\varepsilon$. В этом методе решение уравнения Пуассона, сводится к нахождению поверхностных зарядов, индуцируемых молекулой растворяемого вещества на поверхности, ограничивающей диэлектрик-растворитель. В этом методе вместо решения уравнения Пуассона в трехмерном пространстве надо решить соответствующие уравнения на ограничивающей растворитель поверхности *SES*. Будем считать, что внутри этой поверхности диэлектрическая проницаемость равна $\varepsilon_{in}=1$, а вне неё диэлектрическая проницаемость равна $\varepsilon$. Тогда полярная составляющая энергии взаимодействия субстрата с растворителем определяется выражением:

$$\Delta G_{pol} = \frac{1}{2}\sum_i Q_i \int_{SES} \frac{\sigma(\vec{r})}{\left|\vec{R}_i - \vec{r}\right|} dS, \qquad (1)$$

где $\sigma(\vec{r})$ – плотность поляризационного заряда в точке $\vec{r}$ на поверхности *SES*, $\vec{R}_i$ – вектор, определяющий положение заряда $Q_i$ каждого из атомов молекулы субстрата, интеграл берется по поверхности *SES*, а суммирование производится по всем зарядам атомов субстрата. Плотность поверхностных зарядов $\sigma(\vec{r})$ находится из уравнения PCM, имеющего вид [3, 21, 22]:

$$\sigma(\vec{r}) = \frac{(1-\varepsilon)}{2\pi(1+\varepsilon)}\left(\sum_i \frac{Q_i\left((\vec{r}-\vec{R}_i)\cdot\vec{n}\right)}{\left|\vec{r}-\vec{R}_i\right|^3} + \int_{SES} \frac{\sigma(\vec{r}')\left((\vec{r}-\vec{r}')\cdot\vec{n}\right)}{\left|\vec{r}-\vec{r}'\right|^3} dS'\right). \qquad (2)$$

Интеграл в правой части (2) сингулярен при $\vec{r}=\vec{r}'$ и, при его вычислении вокруг точки $\vec{r}$ выделяется $\delta$-окрестность, которая затем стремится к нулю.

Для численного решения уравнения (2) поверхность *SES* разбивается на малые элементы, и при этом уравнение (2) преобразуются к матричной форме:

$$\mathbf{Aq} = \mathbf{BQ}, \qquad (3)$$

где $\mathbf{A}=\{a_{ij}\}$ – квадратная матрица размера *NxN*, зависящая от параметров поверхностных элементов, *N* – число поверхностных элементов. $\mathbf{B}=\{b_{ij}\}$ – матрица размера *NxM*, зависящая от геометрических параметров поверхностных элементов и положения зарядов $Q_i$ атомов молекулы. *M* – число атомов молекулы, $\mathbf{q}=\{q_i=\sigma_i S_i\}$ – вектор-столбец зарядов поверхностных элементов, $\sigma_i$ – средняя плотность заряда *i*-ого поверхностного элемента, $S_i$ – площадь *i*-ого поверхностного элемента. $\mathbf{Q}=\{Q_i\}$ – вектор столбец зарядов атомов молекулы. При этом система (3) в отличии от системы (2) оперирует с зарядами поверхностных элементов, а не с поверхностной плотностью заряда.

Элементы матрицы **B** имеют вид:

$$b_{ij} = \frac{\vec{n}_i\left(\vec{r}_i - \vec{R}_j\right)}{\left|\vec{r}_i - \vec{R}_j\right|^3} S_i \frac{1-\varepsilon}{4\pi(\varepsilon+1)}. \qquad (4)$$

Здесь введены обозначения: $\vec{r}_i$ - вектор, определяющий положение центра $i$-ого поверхностного элемента, $\vec{n}_i$ - вектор нормали к $i$-ому поверхностному элементу, проведенный через его центр и направленный в сторону диэлектрика-растворителя, $\vec{R}_j$ - вектор, определяющий положение центра $j$-ого атома. С точностью до коэффициента $(1-\varepsilon)/(4\pi(1+\varepsilon))$ каждый элемент матрицы B есть телесный угол, под которым виден $i$-ый поверхностный элемент из $j$-ого атома.

Для суммы элементов каждого столбца матрицы **B** выполняется равенство, которое используется в программе *DISOLV* для коррекции численных ошибок:

$$\sum_{i=1}^{N} b_{ij} = \frac{1-\varepsilon}{1+\varepsilon}, \quad \forall \; j = 1...M \; . \tag{5}$$

Элементы матрицы **A** имеют вид [22]:

$$a_{ij} = \begin{cases} \dfrac{\vec{n}_i(\vec{r}_i - \vec{r}_j)}{|\vec{r}_i - \vec{r}_j|^3} S_i \dfrac{\varepsilon-1}{4\pi(\varepsilon+1)} & i \neq j \\ a_{jj} = \dfrac{\varepsilon}{\varepsilon+1} - \sum_{i \neq j} a_{ij} & i = j \end{cases} \tag{6}$$

В выражении (6) для определения диагональных элементов использовано условие нормировки:

$$\sum_{i} a_{ij} = \frac{\varepsilon}{\varepsilon+1} \quad j = 1...N \; , \tag{7}$$

являющееся следствием равенства [22]:

$$\iint_{SES\;SES} \frac{((\vec{r}-\vec{r}\,')\cdot\vec{n})}{|\vec{r}-\vec{r}\,'|^3} dS'dS = 2\pi \cdot S_{SES}, \quad S_{SES} - \text{полная площадь поверхности}.$$

Полный наведённый на поверхности заряд и суммарный заряд всех атомов молекулы связаны соотношением:

$$\sum_{j=1}^{N} q_j = -\left(1 - \frac{1}{\varepsilon}\right)\sum_{i=1}^{M} Q_i \; . \tag{8}$$

Выражение для энергии (1) может быть записано в виде:

$$\Delta G_{pol} = \frac{1}{2}\mathbf{Q}^T \mathbf{D} \mathbf{q} \; , \tag{9}$$

где выражение для матричных элементов матрицы **D** имеет вид:

$$D_{ij} = \frac{1}{|\vec{R}_i - \vec{r}_j|} \; . \tag{10}$$

2. Для больших значений $\varepsilon \gg 1$, как например в случае воды $\varepsilon=78.5$, можно использовать *модель COSMO (**CO**nductor-like **S**creening **MO**del)* [23]. В этой модели растворитель заменяется континуальным диэлектриком с бесконечной диэлектрической проницаемостью, т.е. $\varepsilon = \infty$. Когда наведенный зарядами субстрата поверхностный заряд найден, соответствующая энергия вычисляется по формуле (1), и полученная величина умножается на корректирующий фактор $C_f$:

$$C_f = \left( \frac{\varepsilon - 1}{\varepsilon + \frac{1}{2}} \right). \qquad (11)$$

Относительная погрешность COSMO имеет величину порядка 1/(2ε). Преимущество COSMO над PCM состоит в том, что в линейном матричном уравнении столбец поверхностных зарядов умножается на симметричную, положительно определенную матрицу, элементы которой не зависят от нормалей поверхностных элементов. Для такой матрицы нахождение энергии и аналитических градиентов можно сделать быстрее и точнее. Уравнение COSMO для поляризационного заряда в интегральной форме записывается в следующем виде [23]:

$$\int\limits_{SES} \frac{\sigma(\vec{r}\,')}{|\vec{r}-\vec{r}\,'|} dS' + \sum_{i=1}^{N} \frac{Q_i}{|\vec{r}-\vec{R}_i|} = 0, \qquad (12)$$

где $\vec{r}$ радиус-вектор любой точки на или вне поверхности.

Это уравнение после дискретизации поверхности *SES*, т.е. разбиение её на конечные элементы, записывается в следующем виде:

$$\mathbf{A}^C \mathbf{q} = -\mathbf{D}^T \mathbf{Q}, \qquad (13)$$

где матрица **D** определена формулой (10), верхним индексом $^T$ обозначается транспонирование, а элементы симметричной матрицы $\mathbf{A}^C$ имеют вид:

$$\begin{cases} a^C_{ij} = \left( \dfrac{1}{|\vec{r}_i - \vec{r}_j|} \right) & i \neq j \\ a^C_{ii} = \dfrac{2\sqrt{3.83}}{\sqrt{S_i}} & , i = j \end{cases} \qquad (14)$$

Заметим, что если ввести матрицу **B** с элементами $b_{ij} = -d_{ji}$, то уравнение (13) примет вид (3) – такой же как и в методе PCM.

Энергия поляризации в матричной форме записывается в виде:

$$\Delta G_{pol} = \frac{1}{2} C_f \mathbf{Q}^T \mathbf{D} \mathbf{q}. \qquad (15)$$

Из теоремы Остроградского-Гаусса можно получить тождество:

$$\sum_{j=1}^{N} q_j = -\sum_{i=1}^{M} Q_i, \qquad (16)$$

где *N* - число поверхностных элементов, *M* - число зарядов внутри полости.

3. Выражение для энергии поляризации в рамках эвристической модели Surface Generalized Born (S-GB) имеет вид [24]:

$$\Delta G_{pol} = -\frac{1}{2} \frac{1}{\left(1+\frac{1}{2\varepsilon}\right)} \cdot \left(1 - \frac{1}{\varepsilon}\right) \cdot \sum_{i,j} \frac{Q_i \cdot Q_j}{\sqrt{|\vec{R}_{i,j}|^2 + a_i \cdot a_j \cdot \exp(-\frac{|\vec{R}_{i,j}|^2}{c \cdot a_i \cdot a_j})}}, \qquad (17)$$

где $\vec{R}_{i,j} = \vec{R}_i - \vec{R}_j$; *c*- эмпирическая константа равная 8, а борновские радиусы $a_i$ находятся через интегралы по поверхности SES:

$$a_i = \frac{1}{2\left(\sum_{n=4}^{7} A_n \cdot I_n^i - A_0\right)}. \tag{18}$$

Здесь $A_n$ – эмпирически константы, определенные в работе [24], $I_n$ – интегралы по поверхности SES:

$$I_n^i = \left[\oint \frac{(\vec{n}_s \cdot (\vec{r}_s - \vec{R}_i))dS}{|\vec{r}_s - \vec{R}_i|^n}\right]^{1/(n-3)} \quad 7 \geq n \geq 4 \;. \tag{19}$$

4. Метод *PCM* с укрупненными поверхностными элементами [22] описывается теми же исходными интегральными уравнениями, что и обычный метод PCM (1), (2), но при дискретизации малые поверхностные элементы объединяются в группы элементов, находящихся наиболее близко к данному поверхностному атому. Будем считать, что поверхностная плотность заряда одинакова во всех точках для любого укрупненного поверхностного элемента. Тогда уравнение PCM с укрупненными поверхностными элементами в матричной форме имеет вид:

$$Rq^{big} = EQ, \tag{20}$$

где $q^{big}$ - столбец зарядов укрупненных поверхностных элементов.

Матричные элементы для укрупненных элементов будут вычисляться по формулам:

$$\begin{cases} R_{jk} = \dfrac{\left(\sum_{l_j}\sum_{m_k} a_{l_j m_k} S_{m_k}\right)}{\sum_{m_k} S_{m_k}} & j \neq k \\ R_{kk} = \left(\dfrac{\varepsilon}{1+\varepsilon}\right) - \sum_{j \neq k} R_{jk} & j = k \;, \\ E_{ji} = \sum_{l_j} b_{l_j i} \end{cases} \tag{21}$$

где $l_j$, $m_j$ – группы малых поверхностных элементов, объединенных в один крупный поверхностный элемент; $S_{l_j}$ - площадь малых поверхностных элементов, $a_{l_j m_k}$, $b_{l_j i}$ - матричные элементы для малых поверхностных элементов определяются выражениями (4) и (6). Для $E_{ji}$ выполняется условие нормировки, аналогичное (5):

$$\sum_j E_{ji} = \frac{1-\varepsilon}{1+\varepsilon}, \forall i = 1,...,M \;. \tag{22}$$

5. Для подсчета неполярных составляющих энергии сольватации $\Delta G_{np} + \Delta G_{cav}$ используется широко применяемая формула [25]:

$$\Delta G_{np} + \Delta G_{cav} = \sigma S_{SAS} + b \;, \tag{23}$$

где $S_{SAS}$ - площадь поверхности типа *SAS* (Solvent Accessible Surface). Отметим, что поверхность доступная растворителю *SAS* образуется центрами молекул растворителя, касающимися молекулы растворяемого вещества (субстрата). В континуальной модели растворителя предполагается, что неэлектростатическая составляющая взаимодействия субстрата с растворителем пропорциональна числу молекул растворителя, соприкасающихся с молекулой субстрата, а коэффициент пропорциональности определяет силу Ван-дер-Ваальсова взаимодействия молекул растворителя с субстратом. В этот же коэффициент и поправочную константу неявно включен также и учёт кавитационной энергии, т.е. свободной энергии необходимой для создания в однородном растворителе полости, в которую помещается молекула субстрата. Поверхность *SAS* по занимаемому объёму больше, чем *SES* для одной и той же молекулы субстрата, и может быть получена из второй соответствующим преобразованием подобия.

В работе [25] для воды используются параметры: *σ=0.00387 ккал/(моль Å²), b=0.698 ккал/моль*.

Поскольку в программе *DISOLV* использовалась $\varepsilon_{in}=1$ (вместо $\varepsilon_{in}=2.24$ в [25]), эти параметры были включены в процесс оптимизации наряду с радиусами атомов (см. раздел Валидация), и в результате были получены практически те же значения параметров неполярной составляющей, что и в работе [25]:

$$\sigma=0.00387 \text{ ккал/(моль}Å^2), b=0.698 \text{ ккал/моль}. \qquad (24)$$

**3. Модуль построения поверхностей SES и SAS.**

В программе используется модифицированный модуль построения молекулярной поверхности TAGSS (Triangulated Adaptive Grid Smooth Surface) [18-20], выполняющий две основные стадии построения поверхности: первая – первичная и вторичная обкатка, вторая – генерирование сетки триангуляции с использованием параметров полученных на первой стадии. При этом молекула представляется совокупностью жестких сфер, центры которых расположены в ядрах составляющих молекулу атомов; радиусы этих сфер различны для атомов разных типов и являются параметрами континуальной модели растворителя. Первичная обкатка - образование молекулярной поверхности путем обкатки молекулы пробной сферой, имитирующей молекулу растворителя. Вокруг молекулы строятся две поверхности: поверхность SAS, описывающаяся положениями центра пробной сферы обкатки, и поверхность SES, состоящая из точек, достигаемых пробной сферой обкатки и расположенных максимально близко к атомам обкатываемой молекулы. (Рис. 1).[16, 26]

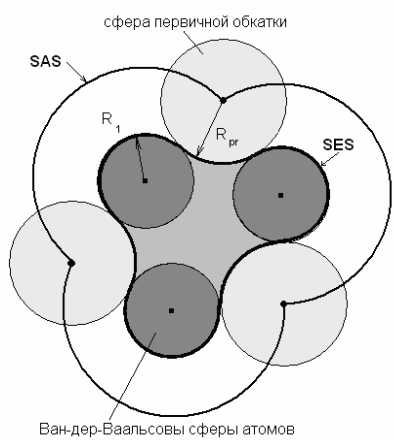

Рис. 1. Первичная обкатка поверхности; $R_{pr}$ – радиус сферы первичной обкатки (радиус молекулы растворителя), $R_1$ – Ван-дер-Ваальсовы радиусы атомов**.**

Процедура первичной обкатки иногда может приводить к нежелательным самопересечениям поверхности и изломам; для их сглаживания применяется процедура вторичной обкатки. Эти нерегулярности можно разделить на два типа:
1) Самопересечение тора первичной обкатки.
2) Пересечение вогнутых сферических фрагментов первичной обкатки – излом.

Диаметр пробной сферы вторичной обкатки выбирается так, чтобы он был меньше диаметра пробной сферы первичной обкатки, но больше определенного критического расстояния, зависящего от средней величины дискретного элемента поверхности.

На Рис. 2, показана процедура сглаживания изломов и удаления области самопересечения при помощи методов вторичной обкатки.

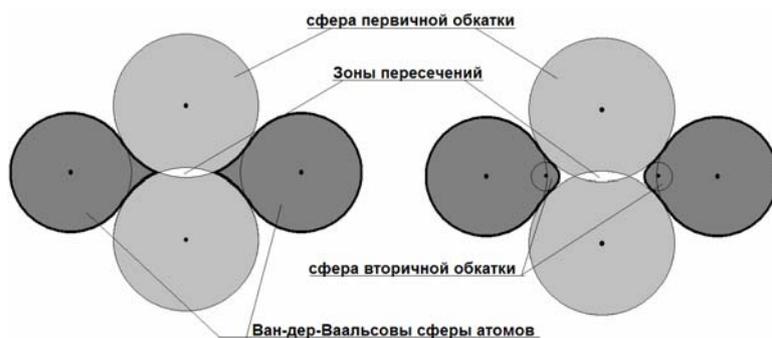

Рис. 2. Вторичная обкатка самопересечения тора первичной обкатки**.**

Стадия обкатки позволяет получить формальное описание поверхности молекулы в виде наборов координат положения и ориентации сферических и тороидальных фрагментов, а также геометрической связанности этих фрагментов друг с другом.

Следующий шаг построения поверхности - генерирование триангуляционной сетки методом последовательного добавления треугольников. Преимущество этого метода – универсальность. Он подходит для любого типа поверхностей, а не только поверхностей, состоящих из сферических и тороидальных сегментов. При этом используется алгоритм проецирования произвольной точки пространства на ближайшую точку поверхности. На основе треугольных поверхностных элементов формируются *многоугольные* поверхностные элементы. Центры многоугольных элементов совпадают с вершинами треугольных элементов. Вершины многоугольных элементов – центры сторон треугольных элементов, выходящие из центров многоугольных элементов и центры тяжести (точка пересечения медиан) этих треугольных элементов.

Поверхность SAS строится из поверхности SES путем преобразования подобия: каждый треугольный элемент на поверхности SAS получается сдвигом трех вершин соответствующего треугольника на поверхности SES на величину радиуса сферы первичной обкатки вдоль нормалей к поверхности SES в этих вершинах. При этом ненулевую площадь будут иметь только те треугольники на SAS, прообразы которых на SES имели хотя бы одну вершину на сферическом сегменте.

**4. Алгоритмы решений.**
1. Уравнение PCM для зарядов поверхностных элементов (3) решалось с помощью следующей итерационной схемы, предложенной в работе [21]:

$$q_k^{(0)} = \frac{\sum_i b_{ki} Q_i}{a_{kk}}, \qquad (25)$$

$$q_k^{(n+1)} = \frac{\left(\sum_i b_{ki} Q_i\right) - \sum_{j \neq k} a_{kj} q_j^{(n)}}{a_{kk}} . \tag{26}$$

Колебания итераций демпфировались путем взятия взвешенной суммы текущей и предыдущей итераций с помощью эмпирического коэффициента:

$$q^{(n+1)}_k = (1-\lambda) q^{(n+1)}_k + \lambda q^{(n)}_k , \quad \lambda = 0.35 . \tag{27}$$

Для подавления численных ошибок использовались тождества (5) и (7), приводящие к выполнению условия (8) для суммы поверхностного заряда.

2. Аналогично решается уравнение РСМ для укрупненных поверхностных элементов

3. Уравнение COSMO для зарядов поверхностных элементов решалось с помощью итерационной схемы метода сопряженных градиентов, используемого для линейных уравнений, определяемых симметричной положительной определенной матрицей [23]. Метод сопряженных элементов [27] минимизирует энергию, описываемую следующей квадратичной формой

$$\Delta G_{pol}(\mathbf{q}) = \frac{1}{2} \mathbf{q}^T \mathbf{A}^C \mathbf{q} - \mathbf{f}^T \mathbf{q} , \tag{28}$$

где $\mathbf{f} = -\mathbf{D}^T \mathbf{Q}$ - столбец правой части уравнения COSMO, $\mathbf{q}$ – столбец зарядов поверхностных элементов. Минимум энергии (28) соответствует решению $\mathbf{q}$ уравнения COSMO.

4. Энергия поляризации в приближении S-GB находится непосредственно из уравнения (17) путем вычисления борновских радиусов всех атомов субстрата по формулам (18) и (19).

**5. Аналитические градиенты.**

Для ускорения оптимизации геометрии молекулы важную роль играет возможность использования аналитических градиентов от энергии молекулы в растворителе. Вычислению градиентов и посвящен настоящий раздел. Очевидно, что точность расчета градиентов должна обеспечить достаточно хорошее совпадение аналитических и численных градиентов при достаточно малом шаге сетки и достаточно малом шаге, который используется при вычислении численных градиентов.

Для удобства введем сопряженные заряды $\mathbf{q}^*$, удовлетворяющие следующему уравнению и в методе РСМ и в методе COSMO:

$$\mathbf{A}^T \mathbf{q}^* = \mathbf{D}^T \mathbf{Q} , \tag{29}$$

где элементы матрицы $\mathbf{D} = \{d_{ij}\}$ определены соотношением (10).

Домножая уравнение (3) слева на матрицу $\mathbf{A}^{-1}$, находим:

$$\mathbf{q} = \mathbf{A}^{-1} \mathbf{B} \mathbf{Q} . \tag{30}$$

Аналогично из уравнения (29) находим

$$\mathbf{q}^* = (\mathbf{A}^T)^{-1} \mathbf{D}^T \mathbf{Q} . \tag{31}$$

Подставляя выражение (30) для $\mathbf{q}$ в выражение (9), получаем в методе РСМ

$$\Delta G_{pol} = \frac{1}{2} \mathbf{Q}^T \mathbf{D} \mathbf{A}^{-1} \mathbf{B} \mathbf{Q} . \tag{32}$$

В методе COSMO в соответствии с выражением (15) в правую часть выражения (32) надо добавить множитель $C_f$ определенный выражением (11).

Обозначим координату $m$-ого атома через $x_m^k$. При этом индекс $k$ пробегает значения 1, 2, 3 и обозначает одну из декартовых координат атома, а индекс $m$ перебирает все номера атомов молекулы. Дифференцируя обе части выражения (32) по координате одного из атомов $x_m^k$ с помощью выражений (30) и (31), получаем выражение для аналитических градиентов в методе PCM через поляризационные $\mathbf{q}$ и сопряженные $\mathbf{q}^*$ заряды и градиенты от матриц $\mathbf{D}$, $\mathbf{A}$ и $\mathbf{B}$.

$$\frac{\partial \Delta G_{pol}}{\partial x_m^k} = \frac{1}{2}\mathbf{Q}^{\mathrm{T}}\frac{\partial \mathbf{D}}{\partial x_m^k}\mathbf{q} - \frac{1}{2}(\mathbf{q}^*)^{\mathrm{T}}\frac{\partial \mathbf{A}}{\partial x_m^k}\mathbf{q} + \frac{1}{2}(\mathbf{q}^*)^{\mathrm{T}}\frac{\partial \mathbf{B}}{\partial x_m^k}\mathbf{Q} \qquad (33)$$

В методе COSMO правая часть выражения (33) домножается на множитель $C_f$, а само выражение может быть записано в еще более компактном виде, учитывая, что в этом методе $\mathbf{B} = -\mathbf{D}^T$, $\mathbf{q}^* = -\mathbf{q}$ и $\mathbf{A} = \mathbf{A}^C$:

$$\frac{\partial \Delta G_{pol}}{\partial x_m^k} = C_f \left( \mathbf{Q}^{\mathrm{T}}\frac{\partial \mathbf{D}}{\partial x_m^k}\mathbf{q} + \frac{1}{2}\mathbf{q}^T\frac{\partial \mathbf{A}^C}{\partial x_m^k}\mathbf{q} \right) \qquad (34)$$

В методе PCM производные матриц $\mathbf{D}$, $\mathbf{A}$, $\mathbf{B}$ находятся из формул (4), (6), (10), а в методе COSMO производные матриц $\mathbf{D}$ и $\mathbf{A}$ находятся из формул (10) и (14). Заметим, что если атомы, по положению которых вычисляется градиент, находятся вблизи поверхности растворителя, то векторы поверхностных элементов $\vec{r}_i$, соответствующие нормали и площади поверхностных элементов, вообще говоря, меняются, и это также надо учитывать при вычислении градиентов.

**6. Аналитические градиенты параметров поверхностных элементов.**

Все поверхностные элементы можно разделить на три основные группы: сферические, тороидальные и смешанные элементы. Сферические элементы – элементы полностью лежащие на одной сфере, тороидальные – полностью лежащие на одном торе, смешанные – пограничные элементы, лежащие одновременно на двух и более фрагментах поверхности. Соответственно, для каждой группы свои правила вычисления производных. Следующие переменные полностью описывают каждый поверхностный элемент: $\vec{r}_i$ – радиус-вектор $i$-ого поверхностного элемента, $\vec{n}_i$ - вектор нормали $i$-ого поверхностного элемента, $S_i$ - площадь $i$-ого поверхностного элемента.

Поверхность меняется только при изменениях координат поверхностных атомов. При этом изменение координат одного поверхностного атома влечёт за собой изменение только небольшой части поверхности – в непосредственной близости от этого атома. В дальнейшем для более компактной записи представлены соотношения между полными дифференциалами параметров поверхностных элементов и полными дифференциалами координат атомов. Из такой записи можно получить выражения и для частных производных. Полные дифференциалы обозначаются знаком $\Delta$, например, $\Delta \vec{r}_i$ означает полный дифференциал радиус-вектора $i$-ого поверхностного элемента. Другие обозначения поясняются по ходу изложения.

1. Сферические элементы.

Изменение координат сферических элементов полностью определяется только смещением центра сферы. При этом изменения нормали и площади поверхностного элемента равны нулю.

$$\Delta \vec{r}_i = \Delta \vec{r}_0 \quad \Delta \vec{n}_i = 0 \quad \Delta S_i = 0, \tag{35}$$

$\Delta \vec{r}_0$ – дифференциал координат центра сферы. Если элемент принадлежит поверхности сферы атома, то смещение центра этой сферы определяется смещением координат атома: $\Delta \vec{r}_0 = \Delta \vec{r}_1$. В других случаях будут более сложные зависимости, которые пояснены ниже.

2. Тороидальные элементы.

Для каждого тороидального фрагмента поверхности вводится локальный ортонормированный базис $\vec{x}, \vec{y}, \vec{z}$ с центром в точке $\vec{p}_c$ (см. рис.3). Ось $\vec{z}$ направлена вдоль прямой соединяющей центры атомов: $\vec{z} = \dfrac{(\vec{r}_2 - \vec{r}_1)}{c}$, $(\vec{x} \cdot \vec{z}) = 0$, $(\vec{y} \cdot \vec{z}) = 0$, $(\vec{x} \cdot \vec{y}) = 0$. Вершина $\vec{p}_0$ определяется с использованием этого локального базиса:

$$\vec{p}_0 = \vec{p}_c + h \cdot \vec{e}_s, \quad \vec{e}_s = \alpha_s \vec{x} + \beta_s \vec{y}, \quad \alpha_s, \beta_s - const, \alpha_s^2 + \beta_s^2 = 1$$

$\alpha_s, \beta_s$ - определяются нормалью точки поверхности $\vec{r}_s$:

$$\gamma_s = \vec{n}_s \cdot \vec{z}, \quad \alpha_s = \dfrac{(\vec{n}_s \cdot \vec{x})}{\sqrt{1 - \gamma_s^2}}, \quad \beta_s = \dfrac{(\vec{n}_s \cdot \vec{y})}{\sqrt{1 - \gamma_s^2}}.$$

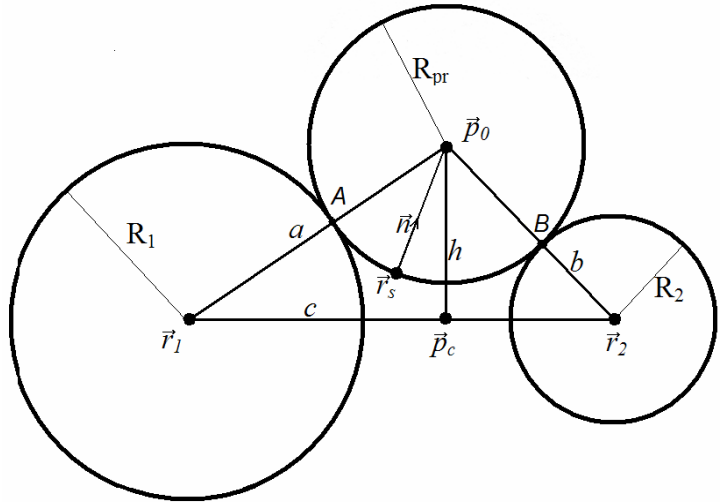

Рис. 3 Геометрическая конфигурация атомов и сферы первичной обкатки при формировании тороидального фрагмента поверхности. Здесь $R_{pr}$ - радиус сферы первичной обкатки, $R_1$ – радиус первого атома, $R_2$ – радиус второго атома; плоскость рисунка задают центры атомов $\vec{r}_1$ и $\vec{r}_2$ и точка поверхности $\vec{r}_s$. Выражения для обозначенных на рисунке величин: $a = R_1 + R_{pr}; b = R_2 + R_{pr}; c = |\vec{r}_2 - \vec{r}_1|$,

$$h = \dfrac{1}{2c}\sqrt{4a^2c^2 - (a^2 + c^2 - b^2)^2}, \quad \vec{p}_c = \tfrac{1}{2}(\vec{r}_1 + \vec{r}_2) + \dfrac{(\vec{r}_2 - \vec{r}_1)(a^2 - b^2)}{2c^2}, \quad \vec{r}_s = \vec{p}_0 - R_{rp}\vec{n}_s.$$

При изменении координат атомов, изменяется положение и ориентация локального базиса, но величины $\alpha_s, \beta_s$ остаются неизменными.

Дифференциалы вспомогательных величин связаны с дифференциалами координат атомов $\Delta \vec{r}_1$ и $\Delta \vec{r}_2$ следующим образом:

$$\Delta c = \frac{(\Delta \vec{r}_2 - \Delta \vec{r}_1) \cdot (\vec{r}_2 - \vec{r}_1)}{c}$$

$$\Delta h = \left(\frac{b^2 + a^2 - c^2}{2h} - h\right) \cdot \frac{\Delta c}{c}$$

$$\Delta \vec{p}_c = \tfrac{1}{2}(\Delta \vec{r}_1 + \Delta \vec{r}_2) + \frac{(\Delta \vec{r}_2 - \Delta \vec{r}_1)(a^2 - b^2)}{2c^2} - (\vec{r}_2 - \vec{r}_1)(a^2 - b^2)\frac{\Delta c}{c^3}$$

$$\Delta \vec{z} = \frac{(\Delta \vec{r}_2 - \Delta \vec{r}_1)}{c} - \frac{(\vec{r}_2 - \vec{r}_1)}{c^2}\Delta c$$

$$\Delta \vec{e}_s = -(\vec{e}_s \cdot \Delta \vec{z})\vec{z}$$

Дифференциалы параметров поверхностного элемента:

$$\Delta \vec{r}_s = \Delta \vec{p}_c + \Delta h \cdot \vec{e}_s + h \cdot \Delta \vec{e}_s - R_{rp} \cdot \Delta \vec{n}_s, \qquad (36)$$

$$\Delta \vec{n}_s = \sqrt{1 - \gamma_s^2}\, \Delta \vec{e}_s + \gamma_s \cdot \Delta \vec{z}, \qquad (37)$$

$$\Delta S_s = \frac{\Delta h}{(h - R_{rp}\sqrt{1 - \gamma_s^2})} \cdot \delta S_s. \qquad (38)$$

3. Смешанные элементы.

Смешанные элементы одновременно включают в себя участки различных фрагментов поверхности. Радиус-вектор $\vec{r}_s$ такого элемента суть радиус-вектор некоторой точки, которая лежит либо на сферической, либо на тороидальной поверхности. В соответствии с этим определяются правила нахождения производной для координат и нормали такого поверхностного элемента. Иначе дело обстоит с площадью этого элемента. Каждый элемент является составным из нескольких смежных плоских треугольников, имеющих общую вершину – центр элемента. Площадь смешанного элемента вычисляется как сумма площадей этих треугольников. Соответственно дифференциал площади элемента вычисляется как сумма дифференциалов площадей каждого треугольника.

Треугольник образован вершинами $\vec{r}_1, \vec{r}_2, \vec{r}_3$, образующие вектора треугольника:

$$\vec{a} = \vec{r}_2 - \vec{r}_1, \quad \vec{b} = \vec{r}_3 - \vec{r}_1$$

Площадь треугольника:

$$\vec{s} = [\vec{a} \times \vec{b}], \quad S = \frac{1}{2}|\vec{s}|,$$

а его дифференциал:

$$\Delta S = \frac{1}{4}\frac{(\Delta \vec{a} \cdot \vec{a})b^2 + (\Delta \vec{b} \cdot \vec{b})a^2 - (\Delta \vec{a} \cdot \vec{b})(\vec{a} \cdot \vec{b}) - (\vec{a} \cdot \vec{b})(\vec{a} \cdot \Delta \vec{b})}{S}, \qquad (39)$$

$$\Delta \vec{a} = \Delta \vec{r}_2 - \Delta \vec{r}_1, \quad \Delta \vec{b} = \Delta \vec{r}_3 - \Delta \vec{r}_1.$$

Дифференциалы координат вершин треугольников определяются типом фрагмента, на котором лежит вершина. Вся молекулярная поверхность состоит только из двух типов фрагментов – сферических и тороидальных. Выше рассмотрены правила дифференцирования параметров поверхностных элементов каждого из типов. Однако каждый из этих типов подразделяется на несколько подвидов. Сферические фрагменты могут быть образованы как Ван-дер-ваальсовыми сферами атомов, так и сферами первичной или вторичной обкатки. Для каждого из подвидов имеется различная связь координат центра сферы с координатами атомов. Аналогичная ситуация имеет место и для тороидальных фрагментов, которые могут образовываться как первичной так и вторичной обкаткой. Для дифференциалов элементов тороидальных фрагментов вторичной обкатки достаточно применить найденные выше формулы (36-38), заменив координаты атомов координатами центров соответствующих сфер первичной обкатки и изменив направление нормали на противоположное.

Сферические фрагменты могут быть представлены как фрагментами первичной, так и фрагментами вторичной обкатки. В случае первичной обкатки таким фрагментом является сферический треугольник образованный сферой обкатки при опоре на три атома. Если координаты атомов $\vec{r}_1, \vec{r}_2, \vec{r}_3$, координаты центра сферы $\vec{r}_0$, то связь между дифференциалами имеет вид:

$$\Delta \vec{r}_0 = \frac{(\Delta \vec{r}_1 \cdot (\vec{r}_1 - \vec{r}_0))}{([(\vec{r}_2 - \vec{r}_0) \times (\vec{r}_3 - \vec{r}_0)] \cdot (\vec{r}_1 - \vec{r}_0))}[(\vec{r}_2 - \vec{r}_0) \times (\vec{r}_3 - \vec{r}_0)] + $$
$$\frac{(\Delta \vec{r}_2 \cdot (\vec{r}_2 - \vec{r}_0))}{([(\vec{r}_1 - \vec{r}_0) \times (\vec{r}_3 - \vec{r}_0)] \cdot (\vec{r}_2 - \vec{r}_0))}[(\vec{r}_1 - \vec{r}_0) \times (\vec{r}_3 - \vec{r}_0)] + \quad (40)$$
$$\frac{(\Delta \vec{r}_3 \cdot (\vec{r}_3 - \vec{r}_0))}{([(\vec{r}_1 - \vec{r}_0) \times (\vec{r}_2 - \vec{r}_0)] \cdot (\vec{r}_3 - \vec{r}_0))}[(\vec{r}_1 - \vec{r}_0) \times (\vec{r}_2 - \vec{r}_0)]$$

В случае вторичной обкатки имеют место сферические фрагменты двух типов. Первый тип – аналогичен сферическим фрагментам первичной обкатки, и для координат его центра применима формула (40), только вместо координат атомов будут фигурировать соответствующие координаты центров сфер первичной обкатки. Второй тип соответствует положениям сферы вторичной обкатки при опоре на сужающуюся горловину тора первичной обкатки (Рис.4).

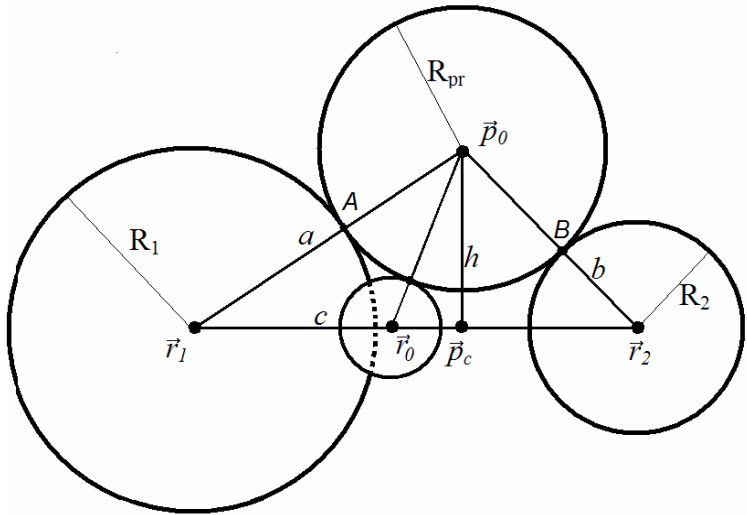

Рис. 4 Схематическое изображение геометрической конфигурации при опоре сферы вторичной обкатки $\vec{r}_0$ на сужающуюся горловину тора первичной обкатки AB.

Координаты центра сферы: $\vec{r}_0 = \vec{p}_c \pm d \cdot \vec{z}$

$d = \sqrt{(R_{pr} + R_{sec})^2 - h^2}$

$R_{sec}$ - радиус сферы вторичной обкатки.

Дифференциалы:

$\Delta d = -\frac{h \cdot \Delta h}{d}$,

$\Delta \vec{r}_0 = \Delta \vec{p}_c \pm (\Delta d \cdot \vec{z} + d \cdot \Delta \vec{z}) \quad (41)$

при этом использованы обозначения в точности такие же, как и в случае с дифференциалами тороидальных элементов, описанные выше.

## 7. Аналитические градиенты энергии SGB.

Полный дифференциал энергии при неизменных зарядах и диэлектрических постоянных ищется прямым дифференцированием выражения для энергии SGB (17):

$$\Delta(\Delta G_{pol}) = -\frac{1}{2}\frac{1}{\left(1+\frac{1}{2\varepsilon}\right)}\cdot\left(1-\frac{1}{\varepsilon}\right)\cdot\sum_{i,j}Q_i\cdot Q_j\cdot\Delta g_{i,j}, \qquad (42)$$

где функция $g_{i,j}$ определяется формулой:

$$g_{i,j}=\frac{1}{\sqrt{\left|\vec{R}_{i,j}\right|^2+a_i\cdot a_j\cdot\exp(-\frac{\left|\vec{R}_{i,j}\right|^2}{c\cdot a_i\cdot a_j})}},$$

$\vec{R}_{i,j}$ - вектор между $i$-ым и $j$-ым атомами. Полный дифференциал:

$$\Delta g_{i,j}=\frac{-g_{i,j}^3}{2}\left[2(\vec{R}_{i,j}\cdot\Delta\vec{R}_{i,j})\left(1-\frac{1}{c}\exp(-\frac{|R_{i,j}|^2}{c\cdot a_i\cdot a_j})\right)+\left(\Delta a_i a_j+a_i\Delta a_j\right)\left(1+\frac{|R_{i,j}|^2}{c\cdot a_i\cdot a_j}\right)\exp(-\frac{|R_{i,j}|^2}{c\cdot a_i\cdot a_j})\right]. \quad(43)$$

Дифференциалы для борновского радиуса $i$-ого атома:

$$\Delta a_i=\frac{-1}{2\left(A_0+\sum_n A_n\cdot I_n^i\right)^2}\cdot\sum_n A_n\cdot\Delta I_n^i \quad 7\geq n\geq 4 \qquad (44)$$

Дифференциалы поверхностных интегралов:

$$\Delta I_n^i=\frac{\Delta J_n^i}{n-3}\left[J_n^i\right]^{\frac{4-n}{n-3}} \quad 7\geq n\geq 4 \text{, где:}$$

$$\Delta J_n^i=\sum_s\Delta J_{s,n}^i, \quad J_{s,n}^i=\frac{(\vec{n}_s\cdot(\vec{r}_s-\vec{r}_i))S_s}{\left|\vec{r}_s-\vec{r}_i\right|^n}, \quad s\text{ – номер поверхностного элемента, }i\text{ - номер атома.}$$

$$\Delta J_{s,n}^i=\left[\frac{[(\vec{n}_s\cdot(\Delta\vec{r}_s-\Delta\vec{r}_i))+(\Delta\vec{n}_s\cdot(\vec{r}_s-\vec{r}_i))]S_s+(\vec{n}_s\cdot(\vec{r}_s-\vec{r}_i))\Delta S_s}{\left|\vec{r}_s-\vec{r}_i\right|^n}-\frac{n(\vec{n}_s\cdot(\vec{r}_s-\vec{r}_i))((\Delta\vec{r}_s-\Delta\vec{r}_i)\cdot(\vec{r}_s-\vec{r}_i))S_s}{\left|\vec{r}_s-\vec{r}_i\right|^{n+2}}\right] \quad(45)$$

Вычисления производных проводятся по формулам (42-45), на основе предварительно вычисленных производных поверхностных элементов (35-41), фигурирующих в (45).

## 8. Градиенты неполярной составляющей энергии сольватации.

Градиент энергии неполярной составляющей:

$$\Delta(\Delta G_{nonpol})=\sigma\sum_{j=1}^{N}\Delta S_j, \qquad (46)$$

где индекс $j=1,...N$ пробегает по всем поверхностным элементам на поверхности SAS. Ненулевой вклад в величины $\Delta S_j$ дают только поверхностные элементы, лежащие вблизи границы сферических фрагментов. Градиенты этих элементов определяются по формулам (35) и (39) поскольку поверхность SAS состоит только из сферических фрагментов.

## 9. Валидация программы.

Все реализованные в программе *DISOLV* методы требуют построения поверхности SES, её триангуляции и решения на ней соответствующих уравнений. При этом модельными параметрами являются: шаг сетки триангуляции и положения её узлов, радиус первичной обкатки, а также параметры, используемые для сглаживания поверхности - радиус вторичной обкатки и критическое расстояние. Прежде всего, валидация должна показать, что можно выбрать такой шаг триангуляционной сетки, что при произвольном её сдвиге по поверхности SES энергия поляризации изменяется достаточно мало – в рамках требуемой для вычисления энергии связывания лиганда с белком точности, составляющей величину в 0.1-1 ккал/моль.

Валидация проводилась при фиксированом значении радиуса сферы первичной обкатки $R_{pr}$=1.4 Å [17, 28] и диэлектрической проницаемости растворителя (воды при комнатной температуре) ε=78.5. При выборе значений остальных модельных параметров были приняты следующие соображения:

- шаг сетки триангуляции должен быть меньше радиусов атомов, на которых строится поверхность, чтобы триангулированная поверхность была достаточно гладкой;
- радиус сферы вторичной обкатки должен быть больше шага сетки триангуляции (чтобы при триангуляции поверхности передавать ее гладкие особенности), но меньше диаметра первичной обкатки и диаметров атомов (чтобы поверхность не сильно деформировалась из-за вторичной обкатки);
- критическое расстояние нужно выбрать существенно меньшим радиуса сферы вторичной обкатки, ибо критическое расстояние является некоторым минимально допустимым критерием кривизны поверхности, при невыполнении которого поверхность сглаживается до значения вторичной обкатки.

Значения радиуса вторичной обкатки и критического расстояния были выбраны 0.4 Å и 0.15 Å, соответственно. Даже сравнительно большие вариации этих величин не приводили к заметному ухудшению точности расчетов (< 0.2 ккал/моль).

На рис. 5 представлены среднее значение и дисперсия полярной части энергии десольватации комплекса белок-лиганд, рассчитанной разными методами, как функции шага сетки триангуляции. Вариации значений энергии десольватации происходили при сдвиге сетки по поверхности SES, задаваемом значениями двух углов, определяющих начальную точку построения сетки, которые равномерно изменялись, принимая по 200 значений в областях своего определения. Усреднение по этим вариациям и давали среднее значение и дисперсию энергии десольватации. Для расчетов был использован комплекс белок-лиганд, взятый из Protein Data Bank [29] – PDB entry 1SQO, содержащий нативный лиганд, состоящий из 34 атомов и имеющий суммарный заряд +1, и белок – уракиназу, состоящий из 3818 атомов и имеющий суммарный заряд +3. В белок программой Reduce [30], а в лиганд программой MolRed [20, 31] были добавлены отсутствующие в PDB атомы водорода, а заряды атомов были расставлены в соответствие с силовым полем MMFF94 [15] с помощью программы докинга SOL [2].

Из рис. 5 можно видеть, что для всех четырех методов расчета при уменьшении шага сетки энергия десольватации выходит на насыщение и становится слабо зависимой как от величины шага, так и от ее положения на поверхности SES. Для трех методов: PCM, COSMO и S-GB ошибка, вносимая вариациями положения триангуляционной сетки на поверхности SES, становится меньше 0.5 ккал/моль уже при шаге сетки 0.3 Å, а при шаге сетки ≤ 0.2 Å становится меньше 0.2 ккал/моль. Для метода укрупненных элементов Абагяна эта ошибка заметно больше и составляет величину 0.5 ккал/моль даже при шаге сетки 0.15 Å.

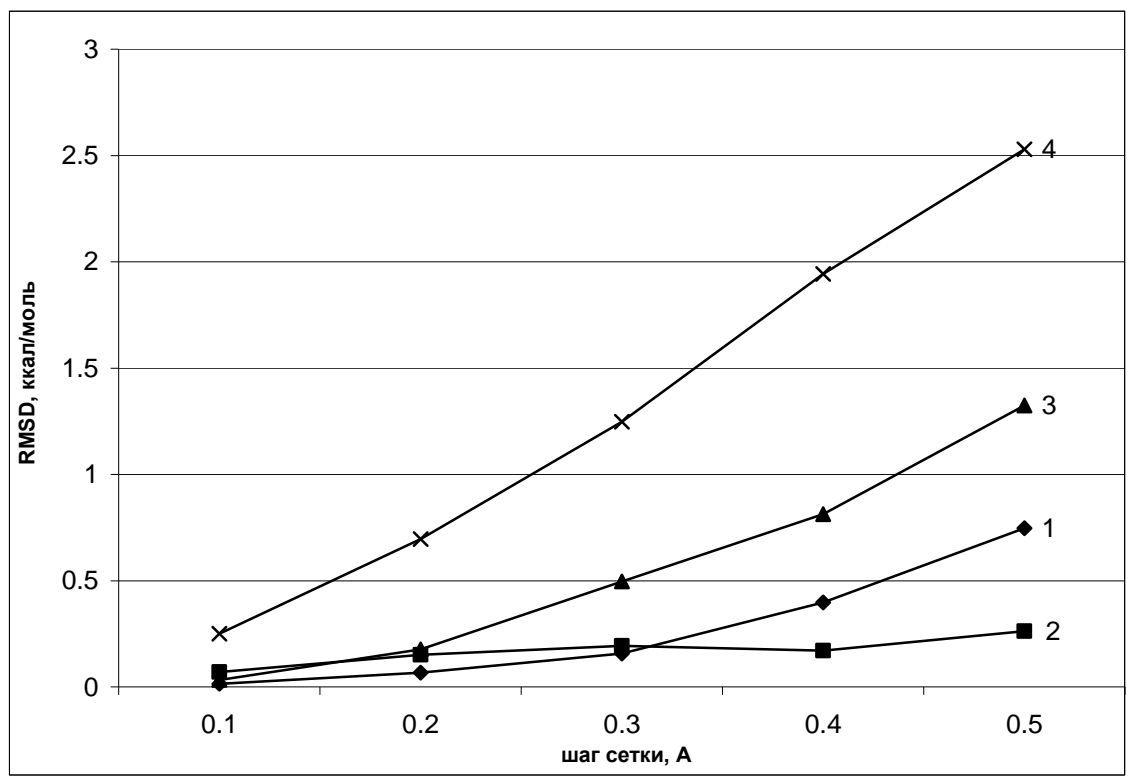

Рис.5. Зависимость среднеквадратичного отклонения RMSD от среднего значения энергии десольватации комплекса урокиназа-лиганд (PDB entry 1SQO) (вследствие изменения ориентации сетки) как функция шага сетки; 1 - SGB, 2 - COSMO, 3 - PCM, 4 - PCM с укрупненными элементами.

Далее, было проведено сравнение энергии десольватации для методов PCM, SGB, COSMO и PCM с укрупненными элементами для двадцати взятых из PDB [32] комплексов при использовании шага сетки 0.2 Å, при котором все эти методы дают ошибку менее 1 ккал/моль для комплекса 1SQO (Рис.5). Комплексы были выбраны достаточно разнообразными, содержащими от 3000 до 10 000 атомов с разбросом зарядов белков и лигандов от -15 до +8 и от -2 до +2, соответственно. Соответствующие структуры были подготовлены также, как это описано выше для урокиназы. Среднеквадратичное отклонение от величины энергии десольватации, посчитанной с помощью наиболее точного метода PCM, составило 4.5, 5.7 и 6.3% для методов COSMO, S-GB и PCM с укрупненными элементами, соответственно.

Правильность вычисления аналитических градиентов проверялась путем их сравнения с численными градиентами. Это было сделано для тромбина, структура которого была взята из базы данных PDB (1O2G) [29] и подготовлена аналогично тому, как описано выше для урокиназы. Для шага сетки 0.1 Å было найдено, что разность величины численных и аналитических градиентов для методов PCM, COSMO и S-GB составляет для поверхностных атомов несколько процентов, а для объемных атомов – на два порядка меньше.

При сравнении рассчитанных величин энергии сольватации молекул с величинами, измеренными экспериментально, важную роль играет выбор атомных радиусов, определяющих поверхность SES. За основу нами были взяты соответствующие радиусы из работы [25]. Однако, в отличие от работы [25] мы не стали использовать ещё один варьируемый параметр – отличную от единицы диэлектрическую проницаемость внутренней области молекул, которая в работе [25] была взята $\varepsilon_{in}$ = 2.21. Мы использовали $\varepsilon_{in}$ = 1 при значении диэлектрической проницаемости воды $\varepsilon$=78.5 и подкорректировали радиусы из работы [25] так, чтобы минимизировать среднеквадратичное отклонение рассчитанных значений энергии сольватации, включая как полярную, PCM, так и неполярную составляющую, от экспериментальных значений для 410 молекул, приведенных в работе [25]. Коэффициенты $\sigma$ и $b$, входящие в формулу (23) для расчета неполярной составляющей энергии сольватации также определялись из условия минимума упомянутого выше среднеквадратичного отклонения. Сравнение полученных рассчитанных и экспериментальных значений энергии сольватации представлены на Рис.6, а найденные оптимальные радиусы атомов представлены в таблице 1 наряду с радиусами из работы [25]. Оптимальные значения коэффициентов $\sigma$ и $b$ составляют 0.00387 ккал/(моль$\text{Å}^2$) и 0.698 ккал/моль, соответственно.

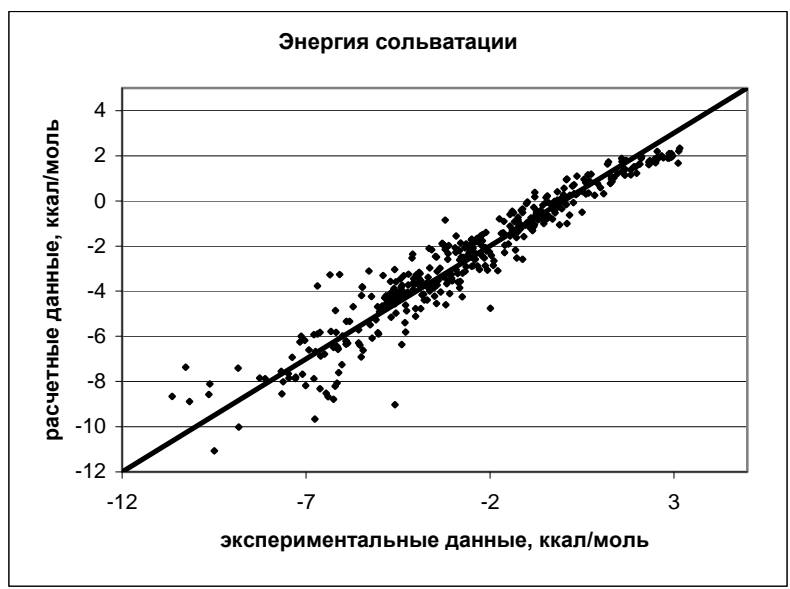

Рис. 6. Сравнение результатов расчета (полярная составляющая - методом PCM) энергии сольватации (точнее, это разность энергий молекулы в вакууме и в воде) с экспериментом для 410 нейтральных молекул из работы [25]. Среднеквадратичное отклонение рассчитанных значений от экспериментальных равно 0.827 ккал/моль.

Таблица 1. Типы атомов силового поля MMFF94 и соответствующие им атомные радиусы из работы [25] и атомные радиусы, найденные в настоящей работе.

| Типы атомов MMFF94 [15] | Атомные радиусы из работы [25], Å | Атомные радиусы, полученные в настоящей работе, Å |
|---|---|---|
| C_1 | 1.46 | 1.95 |
| C_2 | 1.92 | 2.33 |
| C_3 | 2.18 | 2.77 |
| C_4 | 1.84 | 2.02 |
| H_5 | 1.18 | 1.14 |
| O_6 | 1.38 | 1.66 |
| O_7 | 1.35 | 1.49 |
| N_8 | 1.17 | 1.24 |
| F_11 | 2.14 | 2.79 |
| Cl_12 | 1.93 | 2.44 |
| Br_13 | 1.55 | 1.85 |
| I_14 | 1.17 | 1.35 |
| S_15 | 1.62 | 1.96 |
| H_21 | 0.96 | 1.03 |
| C_22 | 1.91 | 2.18 |
| H_23 | 1.34 | 1.71 |
| H_24 | 1.40 | 1.56 |
| H_28 | 0.94 | 1.01 |
| H_29 | 1.01 | 1.07 |
| O_32 | 1.76 | 2.15 |
| C_37 | 1.71 | 2.19 |
| N_38 | 1.70 | 1.96 |
| N_40 | 1.50 | 1.68 |
| N_42 | 1.90 | 2.19 |
| N_45 | 1.87 | 2.16 |

Поскольку не все типы атомов MMFF94 входят в упомянутый выше набор из 410 нейтральных молекул, взятый из работы [25], был образован дополнительный набор из молекулярных ионов, для которых измерены энергии сольватации [33, 34]. Атомные радиусы вновь появившихся типов атомов были получены минимизацией среднеквадратичного отклонения рассчитанных значений энергии сольватации от экспериментальных, при этом найденные ранее атомные радиусы считались неизменными. Сравнение рассчитанных таким образом и экспериментальных значений энергии сольватации представлено на Рис. 7, а полученные дополнительно атомные радиусы и их типы представлены в таблице 2.

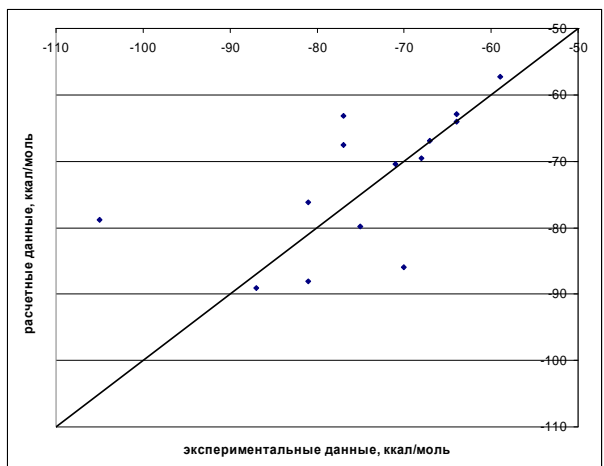

**Рис.7.** Сравнение результатов расчета (полярная составляющая - методом PCM) энергии сольватации с экспериментом для молекулярных ионов из работы [33, 34]. Среднеквадратичное отклонение рассчитанных значений от экспериментальных равно 9.81 ккал/моль.

Таблица 2. Атомные радиусы
для типов атомов, встречающихся
в молекулярных ионах и не представленных
в наборе из 410 нейтральных молекул [25].

| Тип атома MMFF94 | Радиус, Å |
|---|---|
| N_34 | 2.04 |

| O_35 | 1.83 |
|---|---|
| H_36 | 1.25 |
| C_41 | 1.98 |
| O_49 | 2.14 |
| H_50 | 0.88 |
| O_51 | 1.71 |
| H_52 | 0.65 |

Наконец было проведено сравнение результатов расчетов программой DISOLV с аналитически решаемым случаем точечного заряда внутри сферической полости в диэлектрике. При этом энергия поляризации, определяемая как разность электростатической энергии точечного заряда, находящегося в сферической полости диэлектрика, и этого же точечного заряда, находящегося в вакууме, имеет вид:

$$E = -\sum_{n=0}^{\infty}(\varepsilon-1)\frac{Q^2(n+1)r^{2n}}{2(n\varepsilon+\varepsilon+n)a^{2n+1}}, \qquad (47)$$

где $\varepsilon$ – диэлектрическая проницаемость среды, внутри полости диэлектрическая проницаемость равна единице, $a$ – радиус сферической полости, $r$ – расстояние от центра сферы до точечного заряда Q.

Выражение для градиента по нормали к поверхности легко получить дифференцированием выражения (47)

$$\frac{\partial E}{\partial r} = -\sum_{n=1}^{\infty}(\varepsilon-1)\frac{Q^2 n(n+1)r^{2n-1}}{(n\varepsilon+\varepsilon+n)a^{2n+1}} \qquad (48)$$

В программе *DISOLV* такая конфигурация моделировалась двумя фиктивными атомами: один - нейтральный, с радиусом 6 Å, а другой – с единичным зарядом и радиусом 0.5 Å. Результаты расчетов по формулам (51) и (52), а также с помощью программы DISOLV (шаг сетки 0.1 Å) для различных положений единичного заряда приведены в таблице 3.

Таблица 3. Энергия поляризации и её градиенты для точечного единичного заряда, находящегося внутри сферической полости радиуса 6 Å в диэлектрике с диэлектрической проницаемостью $\varepsilon$ =78.5, для различных положений этого заряда; $r$ – расстояние от заряда до центра сферической полости.

| $r$, Å | энергия, ккал/моль | | градиент, ккал/моль·Å | | |
|---|---|---|---|---|---|
| | Выражение (47) | Disolv | Выражение (48) | Disolv, анлитический | Disolv, численный |
| 0 | -27.3201 | -27.3201 | 0 | 1.83E-05 | 2.84E-05 |
| 0.2 | -27.3503 | -27.3503 | -0.302306 | -0.30257 | -0.30262 |
| 3 | -36.3637 | -36.3632 | -8.03456 | -8.03357 | -8.03618 |
| 5 | -88.8714 | -88.8452 | -80.4738 | -81.45514 | -74.56199 |

Как видим, результаты расчетов по формулам (47) и (48) с высокой точностью воспроизводятся программой DISOLV, метод PCM.

Время расчета $t$ энергии сольватации для белков из нескольких тысяч атомов без расчета градиентов зависит от шага сетки $h$ для методов PCM и COSMO как $t \sim h^{-4}$, а для методов S-GB и PCM с укрупненными элементами как $t \sim h^{-2.5}$ и $t \sim h^{-3}$, соответственно. Например, при вычислении сольватации белка урокиназы из комплекса PDB 1SQO для шага 0.2 Å самый точный из использованных методов, PCM, требует на 1 CPU (Intel Xeon E5472, 3.0 ГГц) около 3000 секунд.

Сравнение результатов расчета энергии сольватации белка программой *DISOLV* методом PCM и программой APBS [14] при использовании одинаковых радиусов и зарядов атомов, шага сетки и точности вычислений показало различие в единицы процентов при соизмеримых временах счета.

**10. Заключение.**

В настоящей работе рассмотрен один из аспектов повышения точности компьютерного предсказания ингибирующей активности молекул-кандидатов в ингибиторы для заданных белков-мишеней, а именно учет влияния растворителя на энергию связывания белок-лиганд. В настоящей работе детально описаны физические основы четырех неявных (континуальных) моделей растворителя PCM, COSMO, S-GB и PCM с укрупненными элементами, реализованными в программе DISOLV; приведены методы решения соответствующих уравнений, основы алгоритма построения молекулярных поверхностей, используемых в этих моделях и построения градиентов энергии молекулы в растворителе. Последние нужны при использовании программы DISOLV для локальной оптимизации энергии молекулы в растворителе. Представлены результаты валидации программы DISOLV, которые показали не только возможность достижений хорошей точности расчетов (при произвольных сдвигах триангуляционной сетки) лучше нескольких десятых ккал/моль при разумных временах для таких больших макромолекул как белки, но и неплохое совпадение (среднеквадратичное отклонение 0.8 ккал/моль) рассчитанных значений энергии перехода молекулы из газа в воду с экспериментальными значениями для нескольких сотен нейтральных молекул. Для молекулярных ионов среднеквадратичное отклонение рассчитанных значений энергии сольватации от экспериментальных значительно выше (10 ккал/моль), но и это значение лежит в большинстве случаев в пределах ошибки измерений. В целом результаты валидации показали, что программа DISOLV может быть использована в режиме поспроцессинга для уточнения оценки энергии связывания белок-лиганд, даваемой программой докинга.